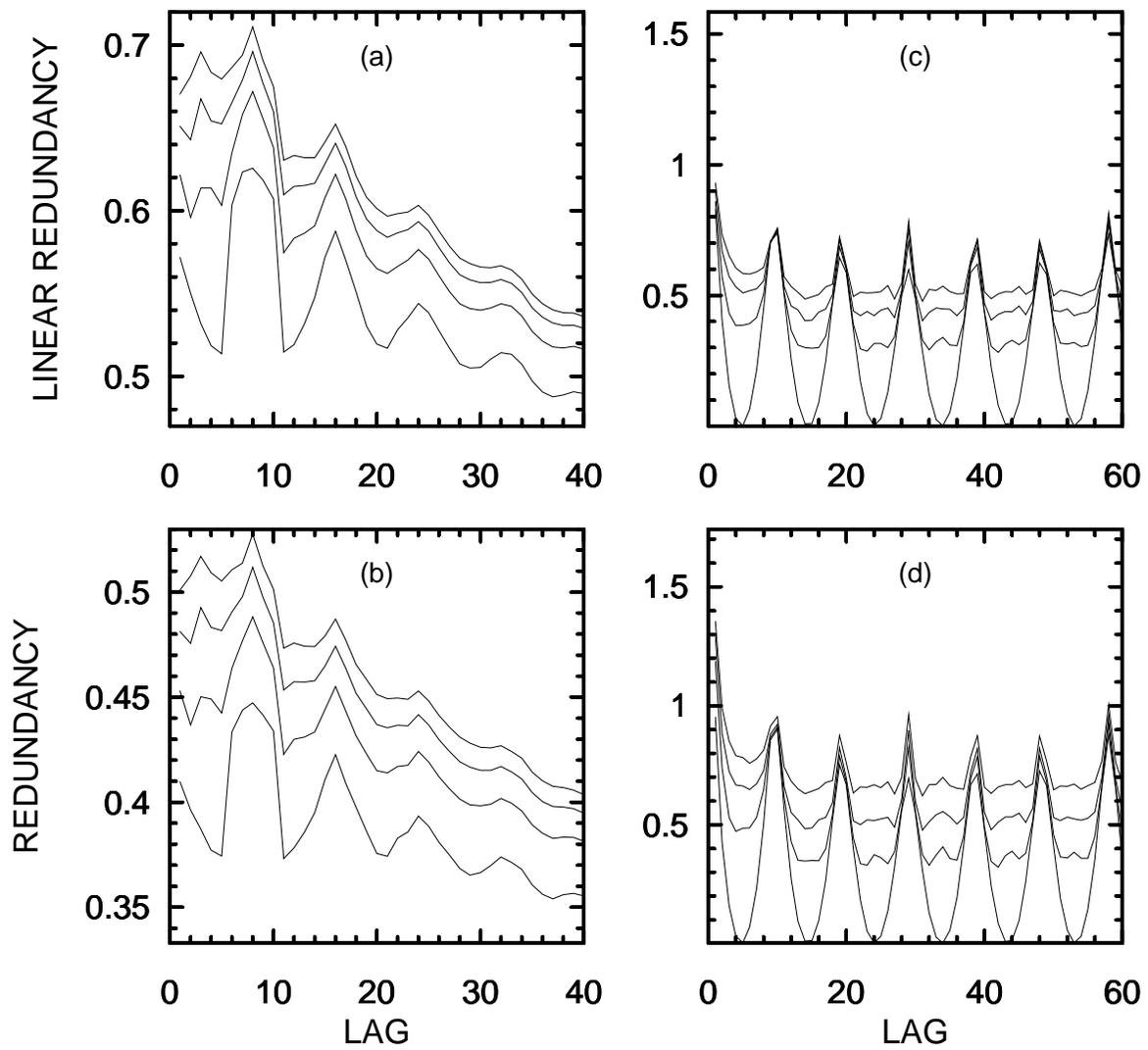

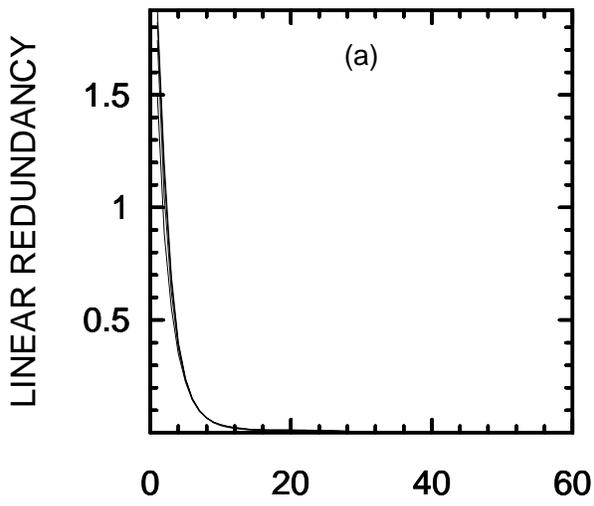
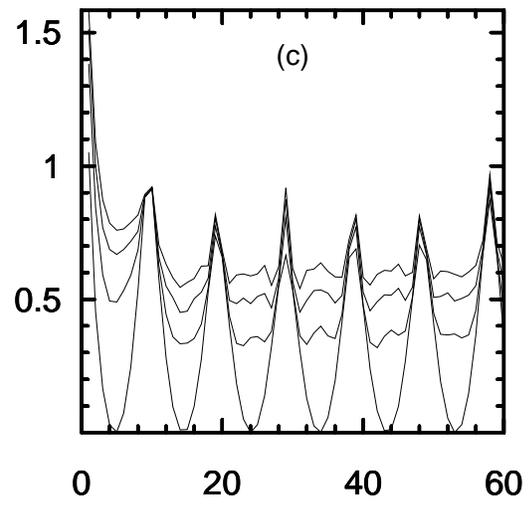
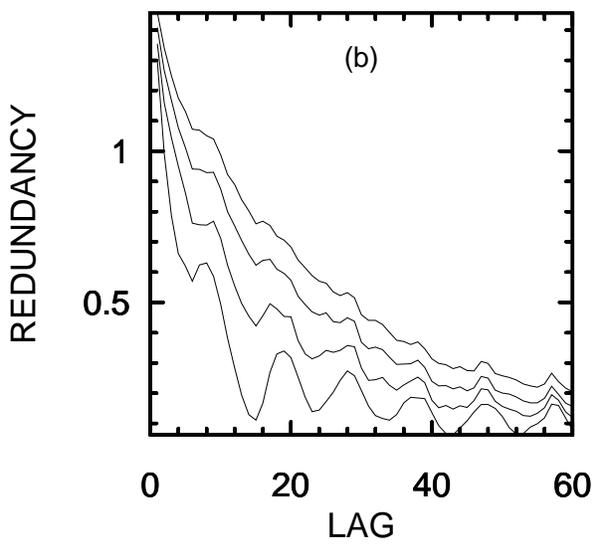
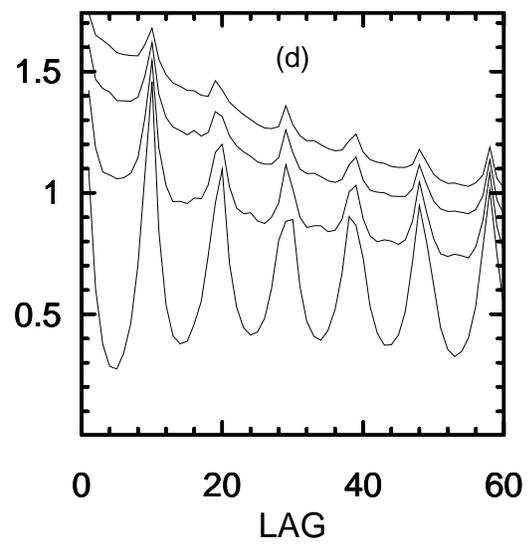

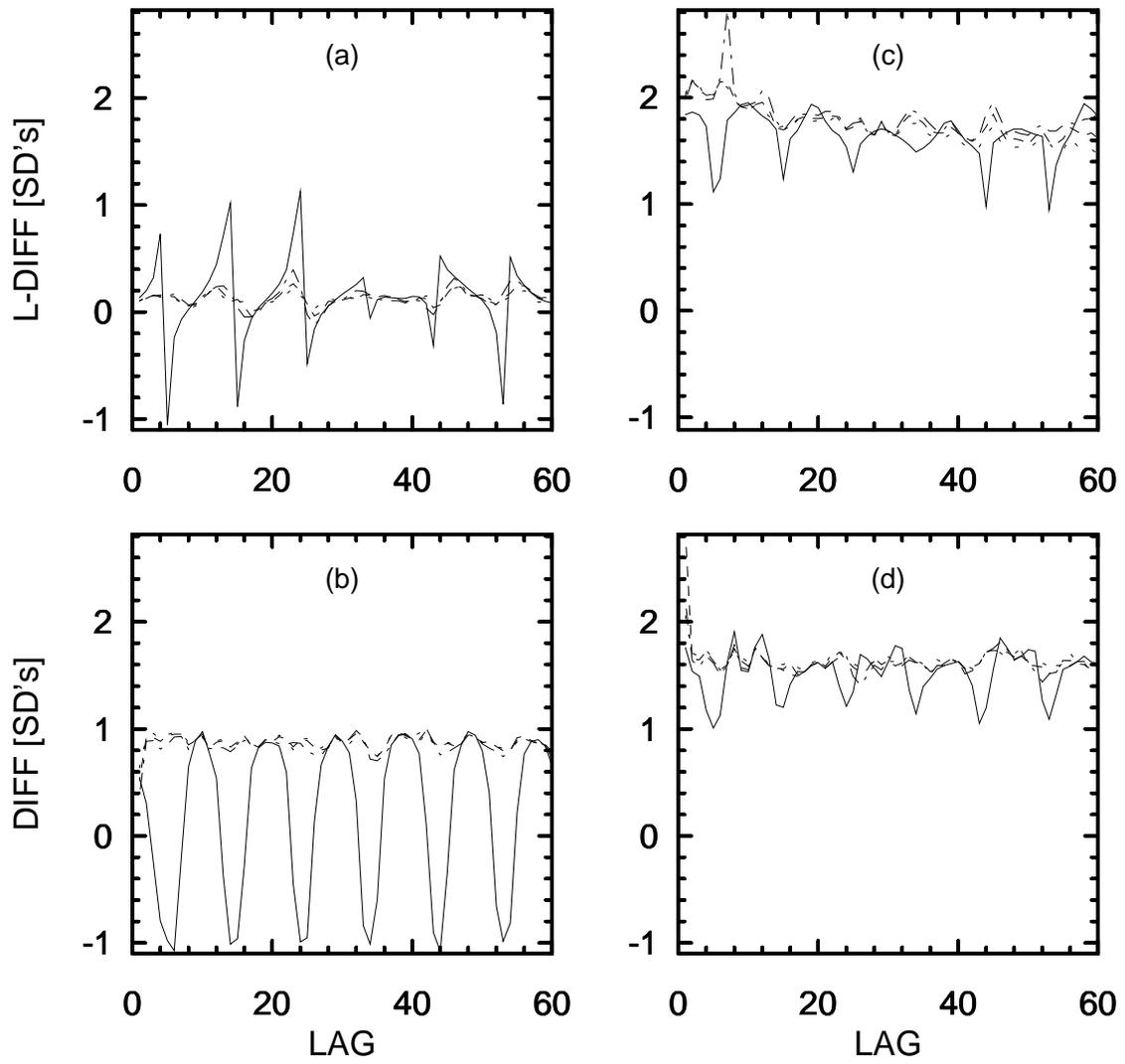

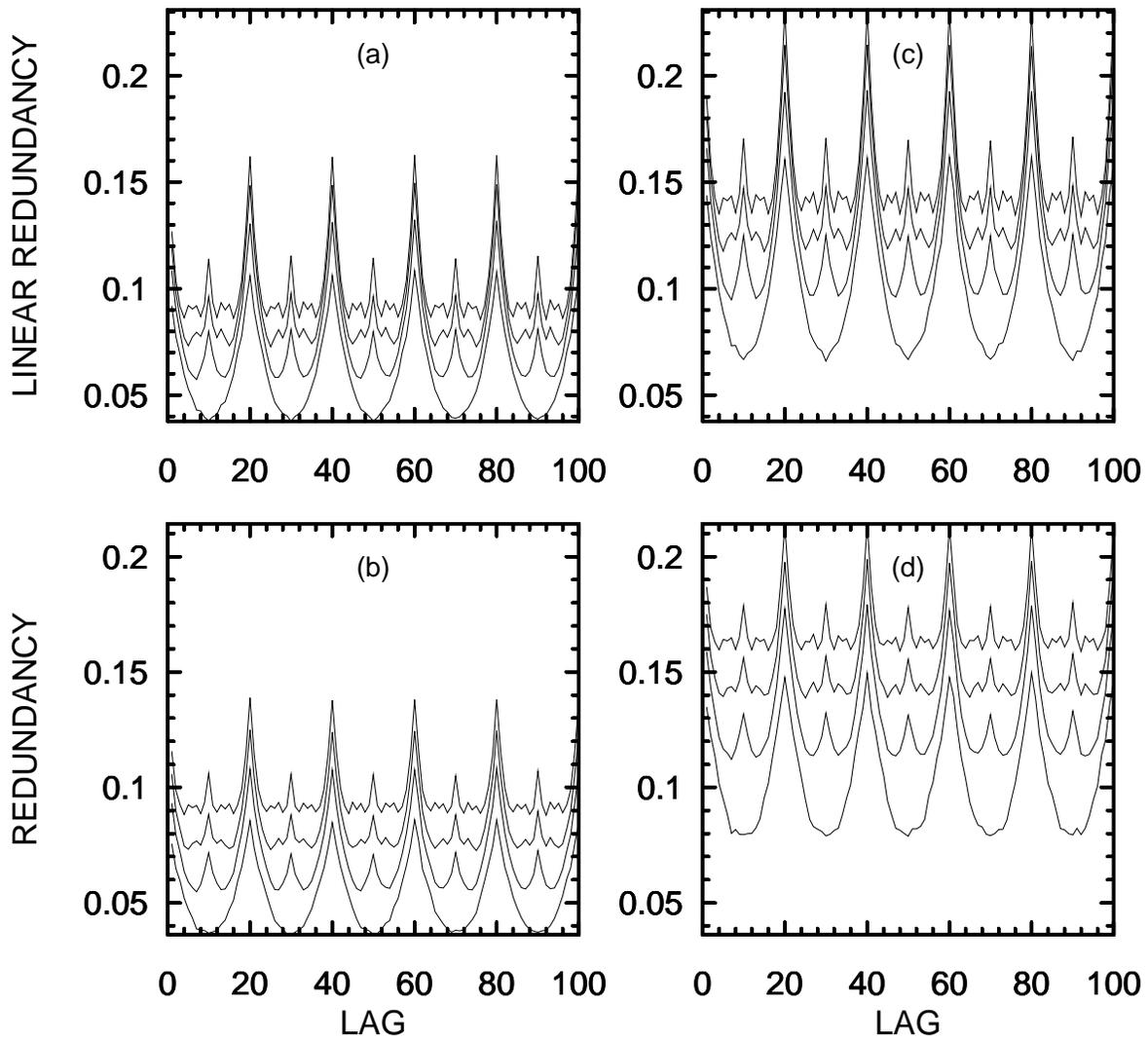

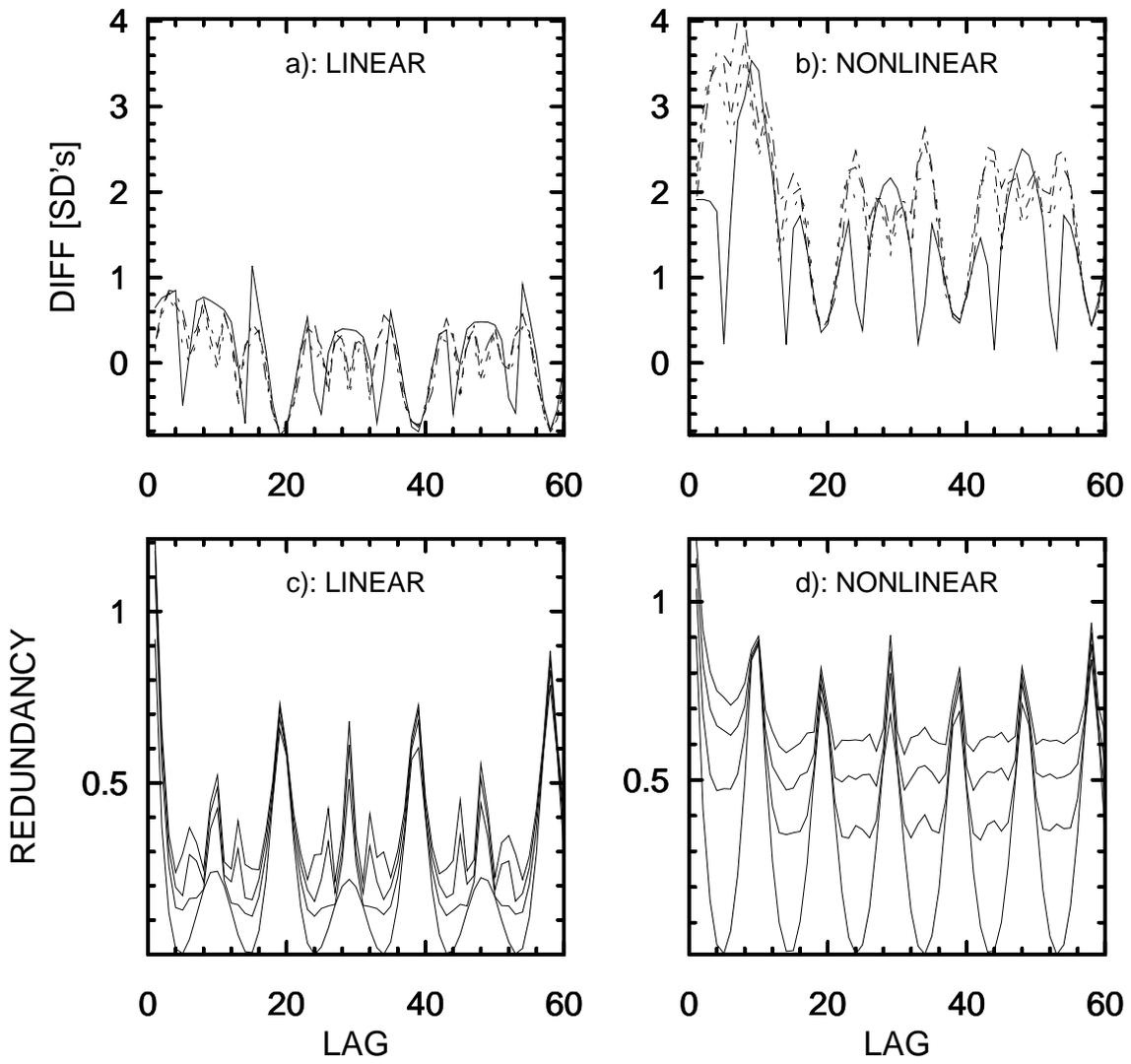

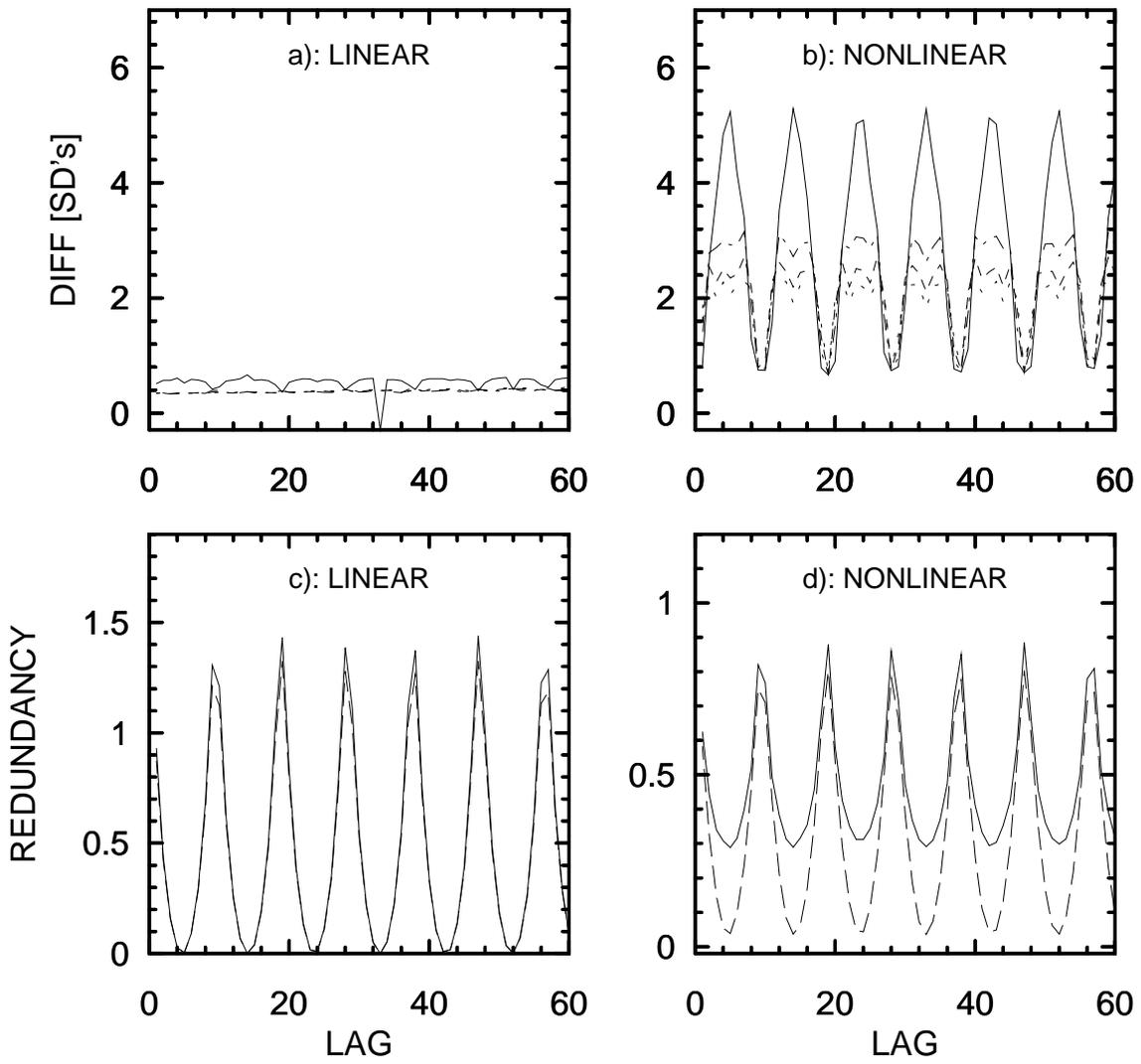

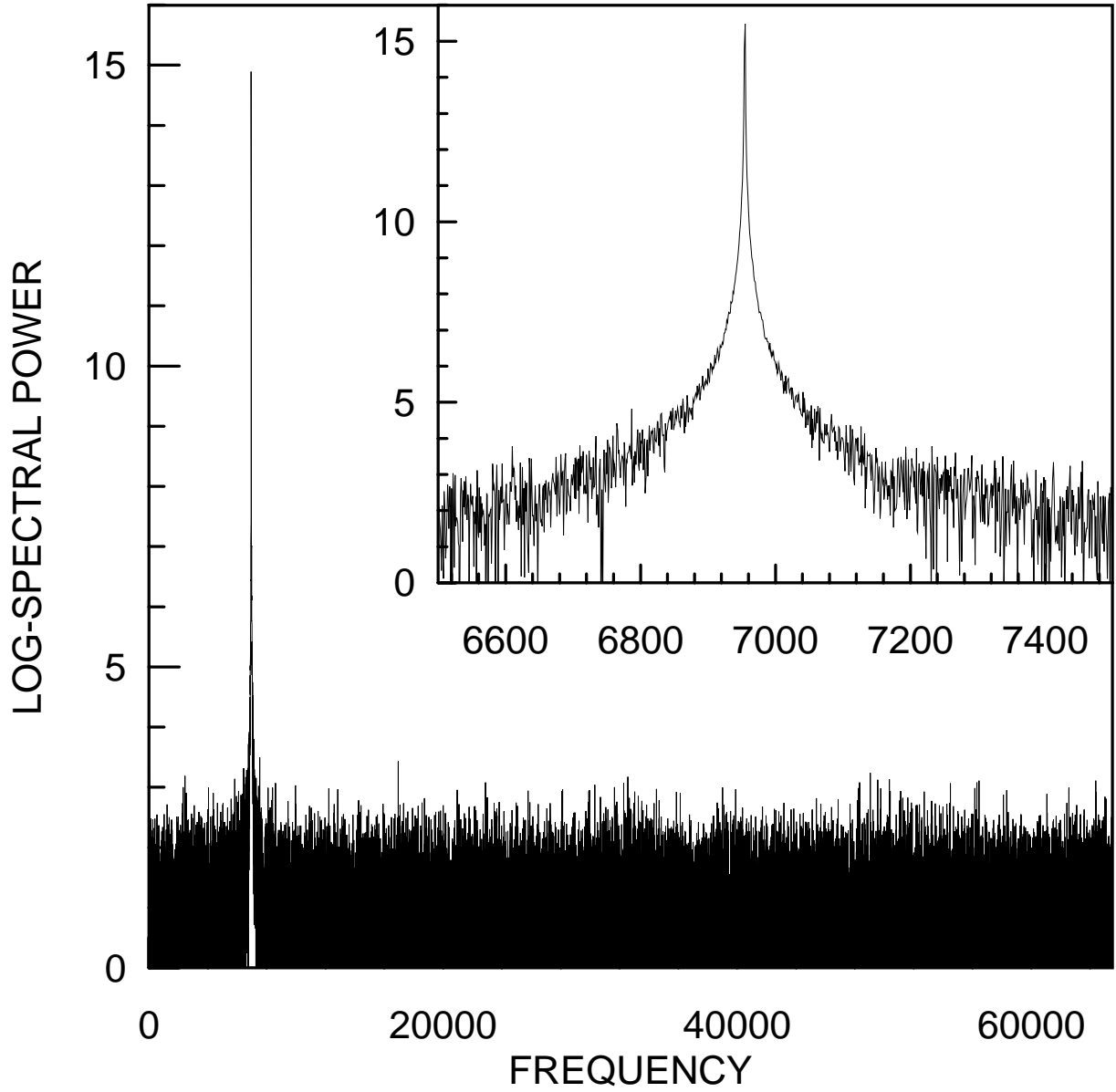

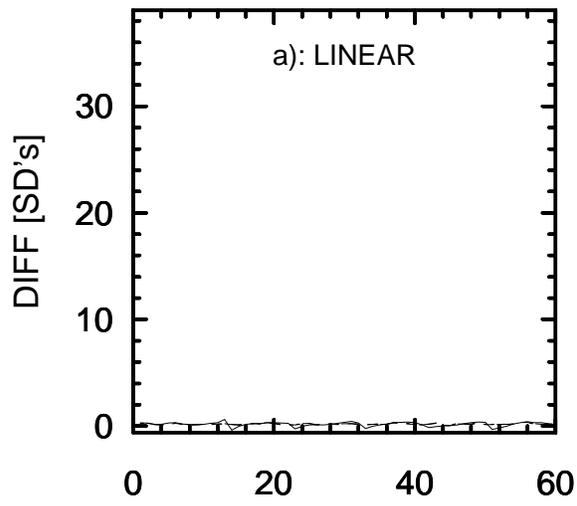
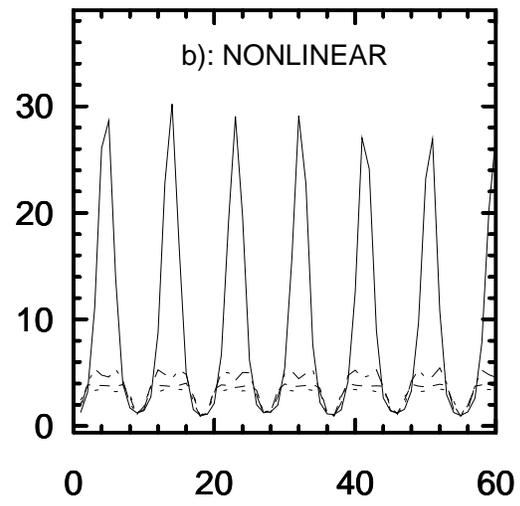
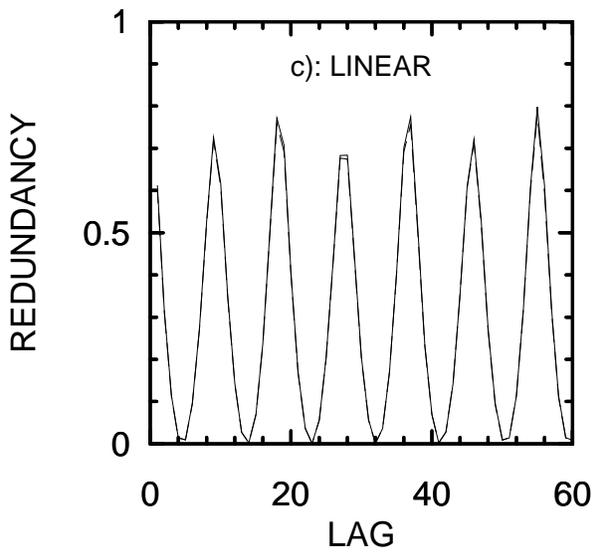
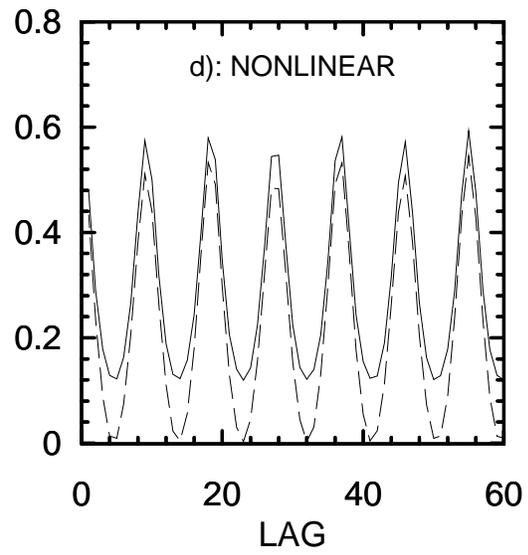

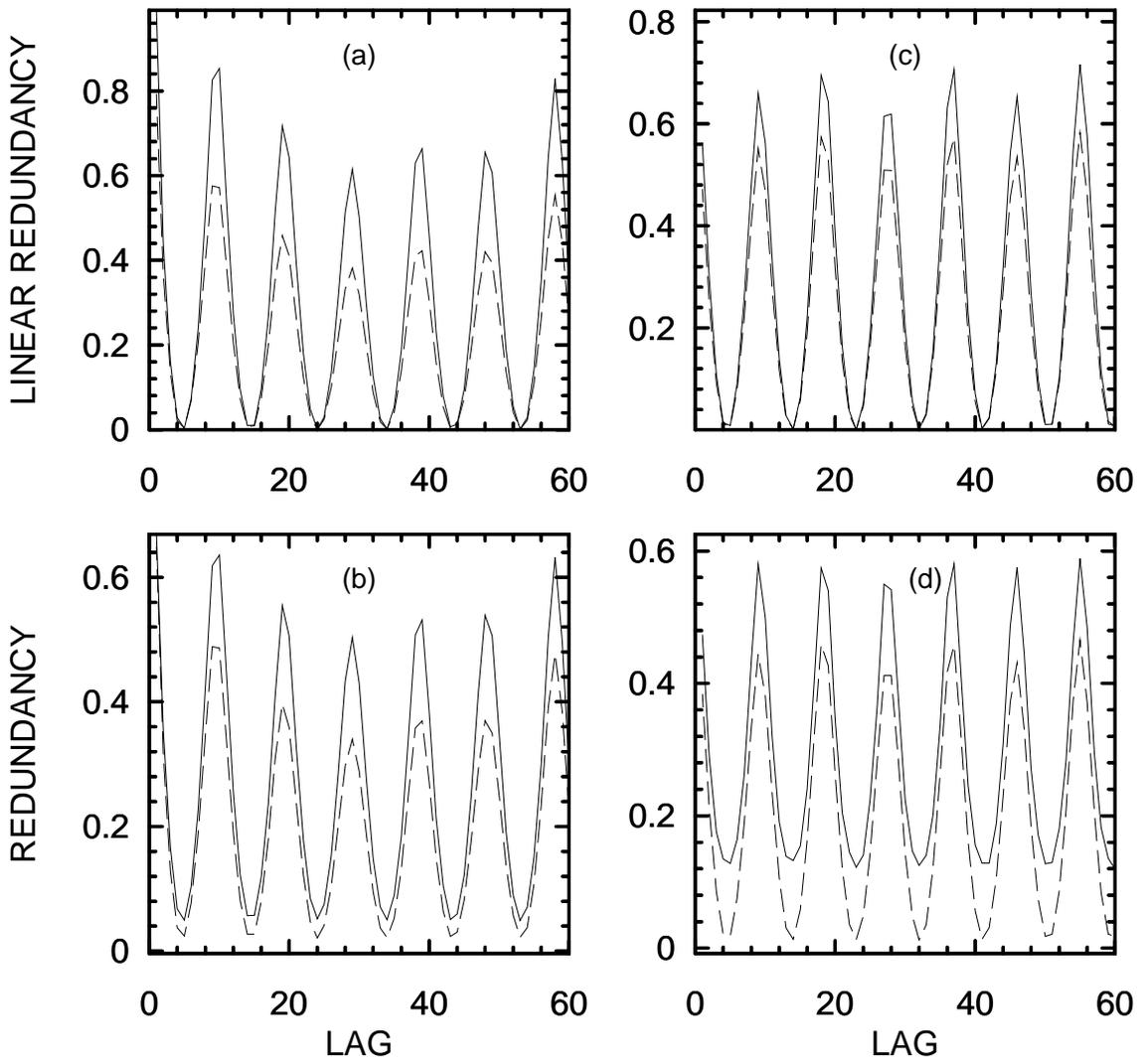

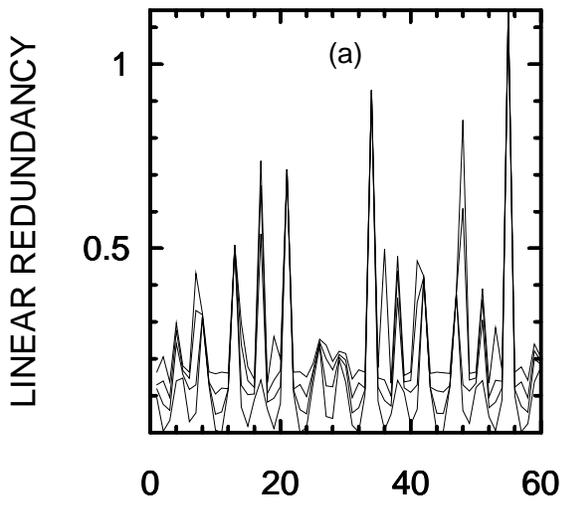
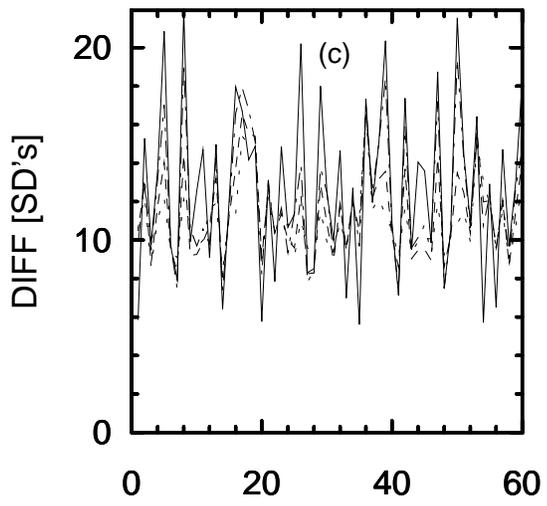
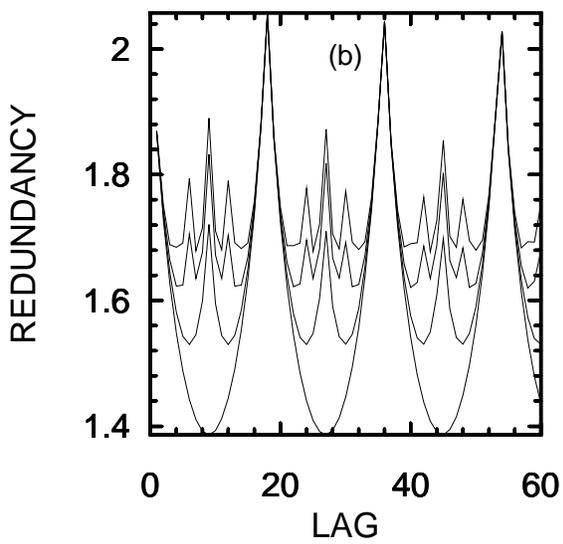
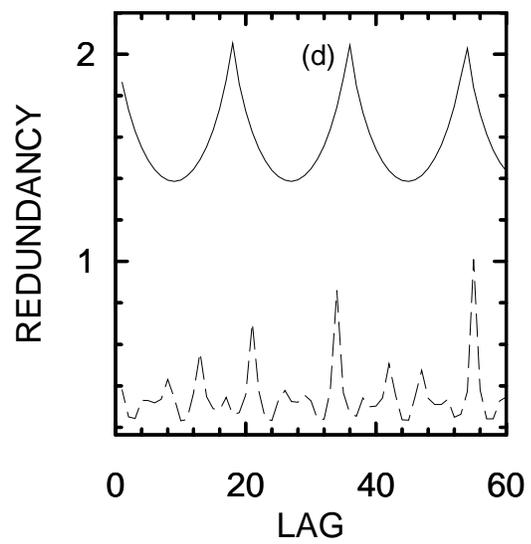

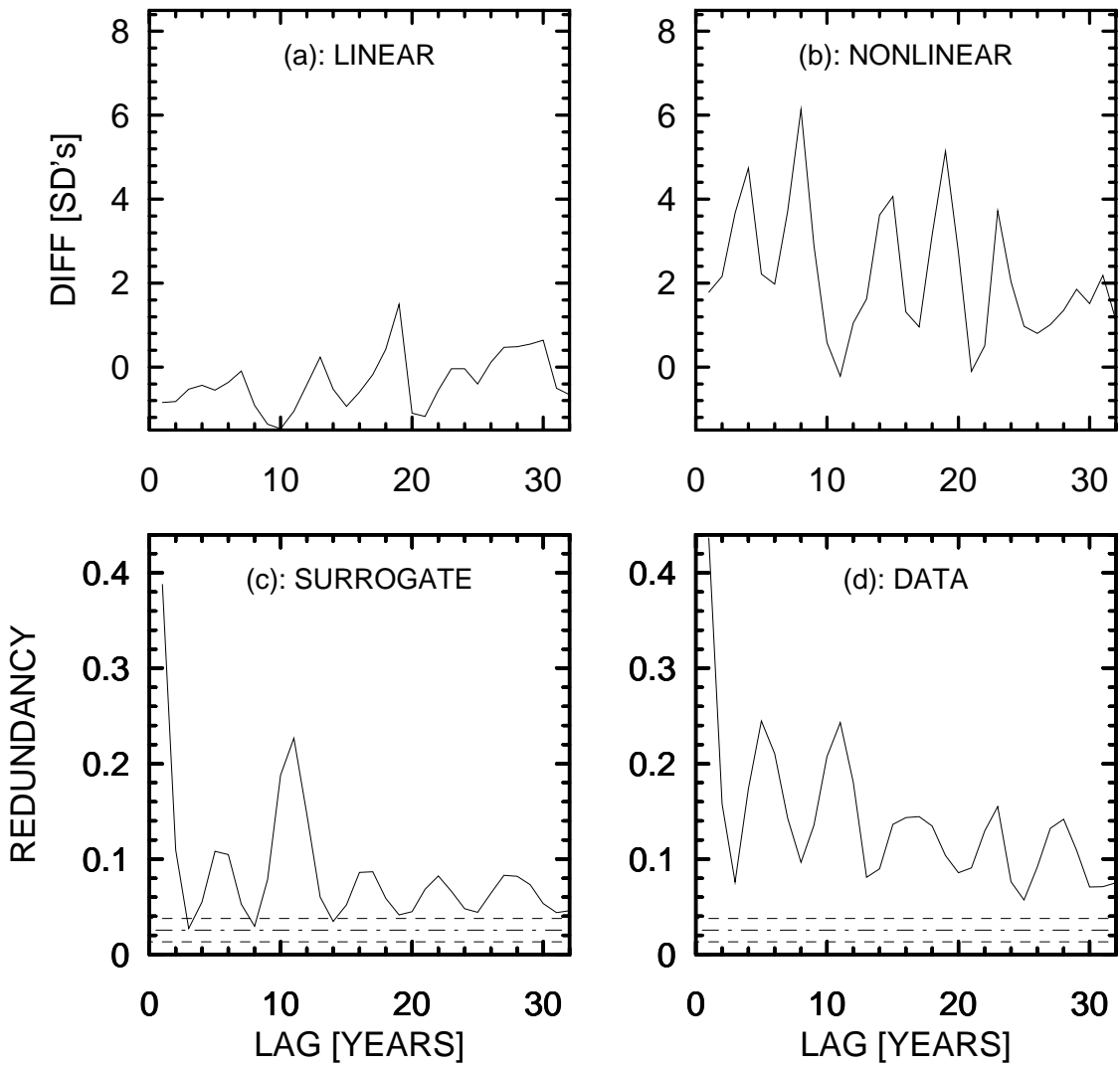

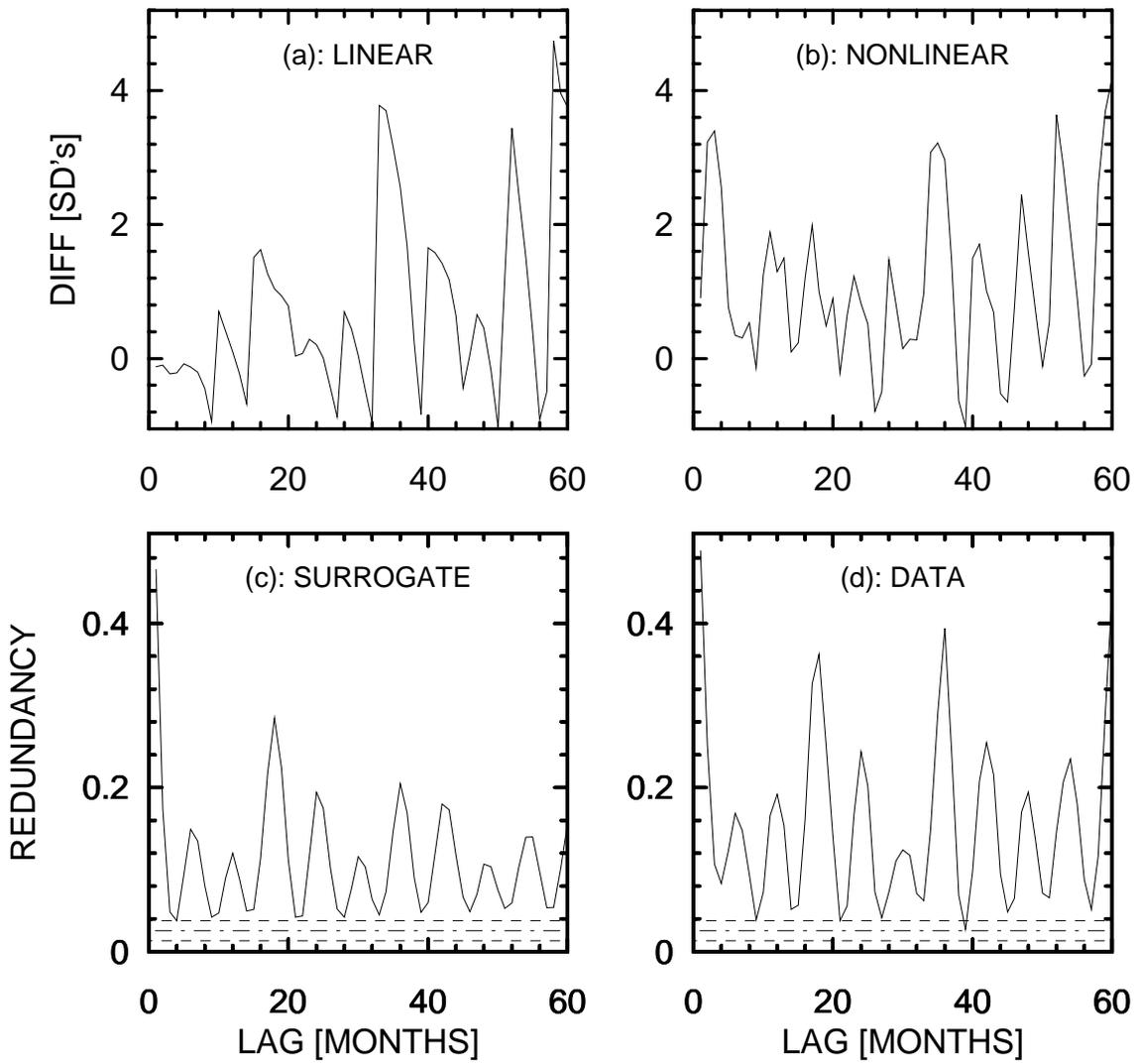

# TESTING FOR NONLINEARITY USING REDUNDANCIES:
## Quantitative and Qualitative Aspects

Milan Paluš
*Santa Fe Institute,*
*1660 Old Pecos Trail, Suite A,*
*Santa Fe, NM 87505, USA*
*E-mail: mp@santafe.edu*

**Abstract**

A method for testing nonlinearity in time series is described based on information-theoretic functionals – redundancies, linear and nonlinear forms of which allow either qualitative, or, after incorporating the surrogate data technique, quantitative evaluation of dynamical properties of scrutinized data. An interplay of quantitative and qualitative testing on both the linear and nonlinear levels is analyzed and robustness of this combined approach against spurious nonlinearity detection is demonstrated. Evaluation of redundancies and redundancy-based statistics as functions of time lag and embedding dimension can further enhance insight into dynamics of a system under study.

## 1 Introduction

The problem of inferring the dynamics of a system from measured data is a perpetual challenge for time series analysts. Ideas and concepts from nonlinear dynamics and theory of deterministic chaos have led to a number of algorithms, able, in principle, to identify and quantify underlying nonlinear deterministic/chaotic dynamics [1, 16, 17, 27]. After extensive use, however, many of these algorithms were found chronically unreliable, often producing spurious dimension or Lyapunov exponent estimates [28, 41, 47, 51, 7] and thus supporting false identification of chaotic dynamics in data, consistent with a simpler explanation [34, 40, 52, 51]. Some authors, considering these problems, proposed to test necessary conditions for chaos like nonlinearity or nonlinear determinism [53, 35, 21]. As it was pointed out by Theiler et al. [53], detection of nonlinearity is considerably easier goal than that of positive identification of chaotic dynamics. On the other hand, detection of nonlinearity is not a trivial task either and there are various sources of possible errors in nonlinearity testing.

In this paper, we demonstrate how the information-theoretic functionals – redundancies can be used for detecting nonlinearity in time series. Besides the general redundancy, we can define its linear version, sensitive to linear dependence only. Comparison of the linear and general redundancies can be used for a "qualitative" test for nonlinearity, as was proposed in [35]. A rigorous statistical quantitative test can be developed by incorporating the technique of surrogate data [53]. The latter, however, can be a source of spurious identification of nonlinearity, as in some cases surrogate data do not fulfill theoretical expectations, like preservation of linear properties of the original data. If this is the case, the pair of linear and general redundancies can be used for testing differences between the data and its surrogates on both the linear and nonlinear levels, and thus prevent spurious results due to changes in the linear properties. The combination of both the quantitative and qualitative testing can further improve reliability of the results.

Some remarks on linearity and nonlinearity are given in Sec. 2. The information theoretic functionals – redundancies are introduced in Sec. 3. Qualitative testing for nonlinearity is presented in Sec. 4. The quantitative method, based on surrogate data technique, is described in Sec. 5. In Sec. 6 numerical problems which can lead to spurious detection of nonlinearity are discussed. Differences and possible interplay of qualitative and quantitative testings are discussed in Sec. 7. Application of the proposed methodology to real data is presented in Sec. 8. The conclusion is given in Sec. 9. In Appendix A numerical procedures for estimating the redundancies are discussed. Numerically generated data, analyzed in this paper, are described in Appendix B.

## 2  Some remarks on linearity and nonlinearity

This paper deals with the problem of detecting nonlinearity in time series, therefore, it seems reasonable to say when a time series $\{y(t)\}$ is nonlinear. Intuitively, we can say that it is the case when the relation between $y(t)$ and $y(t+\tau)$ is nonlinear, or, cannot be expressed by a linear function. Looking for a more rigorous definition, we use the way we will detect nonlinearity here — by showing that a time series is inconsistent with a linear stochastic process. Thus, for detection of nonlinearity we do not need to exhibit or describe the underlying nonlinear dynamics, but simply to find arguments that a linear model is inadequate. This approach – testing data against a null hypothesis – was advocated by a number of authors [11, 8, 9, 10, 54]. Therefore we need to define a linear stochastic processes.

Let $\{y(t)\}$ be a time-series realization of a stationary stochastic process $\{Y(t)\}$. Without loss of generality we can set its mean to zero.

$\{Y(t)\}$ is a stochastic linear process if $Y(t)$ can be written as:

$$Y(t) = Y(0) + \sum_{i=1}^{\infty} a(i)Y(t-i) + \sum_{i=0}^{\infty} b(i)N(t-i), \qquad (1)$$

where $b(0) = 1$, $\sum_{i=1}^{\infty} |a(i)| < \infty$, $\sum_{i=0}^{\infty} |b(i)| < \infty$, $\{N(t)\}$ is an independent, identically distributed (iid), normally distributed process with zero mean and finite variance. (For more details see [39].) The requirement for $\{N(t)\}$ to be an iid process is an important condition, since any stationary process with zero mean posses a Wold decomposition [2] of the form (1), when the process $\{N(t)\}$ is uncorrelated. The principal difference lies in the fact, that while "independent" means also "uncorrelated", "uncorrelated" means only "linearly independent" and not independent in general. This distinction will be demonstrated in Sec. 4.

Now, having in mind the null hypothesis of a linear stochastic process, we will test whether all the dependences in a time series can be detected on the linear level (Sec. 4) and evaluate a quantitative difference between the data under study and a set of realizations of a linear stochastic process with the same linear properties as the data (Sec. 5). If we find evidence that the time series is not consistent with the null hypothesis (of a linear stochastic process), we will consider the series to be nonlinear.

## 3 Mutual information and redundancies

In this section we will define basic functionals from information theory, more details can be found in any book on information theory [4, 15, 24, 25, 45, 6].

Let $X, Y$ be random variables with probability distribution densities $p_X(x)$ and $p_Y(y)$. The entropy of the distribution of a single variable, say $X$, is defined as:

$$H(X) := -\int p_X(x) \log(p_X(x)) dx. \qquad (2)$$

(We use the mark ":=" in order to distinguish definitions from other types of equations.) For the joint distribution $p_{X,Y}(x,y)$ of $X$ and $Y$ the joint entropy is defined as:

$$H(X,Y) := -\int\int p_{X,Y}(x,y) \log(p_{X,Y}(x,y)) dxdy. \qquad (3)$$

The average amount of information, that the variable $X$ contains about the variable $Y$, is quantified by the mutual information $I(X;Y)$:

$$I(X;Y) := H(X) + H(Y) - H(X,Y). \tag{4}$$

Clearly, $I(X;Y) = 0$ iff $p_{X,Y}(x,y) = p_X(x)p_Y(y)$, i.e., iff $X$ and $Y$ are statistically independent.

For $n$ variables $X_1, \ldots, X_n$ the extension of (3) is:

$$H(X_1, \ldots, X_n) :=$$
$$- \int \ldots \int p_{X_1, \ldots, X_n}(x_1, \ldots, x_n) \log(p_{X_1, \ldots, X_n}(x_1, \ldots, x_n)) dx_1 \ldots dx_n. \tag{5}$$

The $n$-dimensional extension of (4) is called *redundancy* and is given by:

$$R(X_1; \ldots; X_n) := H(X_1) + \ldots + H(X_n) - H(X_1, \ldots, X_n). \tag{6}$$

The redundancy of the $n$-tuple $X_1, \ldots, X_n$ vanishes iff there is no dependence among these variables.

Now, let $X_1, \ldots, X_n$ be an $n$-dimensional normally distributed random variable with zero mean and covariance matrix $\mathbf{C}$. In this special case the redundancy (6) can be computed straightforwardly from the definition:

$$R(X_1; \ldots; X_n) = \frac{1}{2} \sum_{i=1}^{n} \log(c_{ii}) - \frac{1}{2} \sum_{i=1}^{n} \log(\sigma_i), \tag{7}$$

where $c_{ii}$ are the diagonal elements (variances) and $\sigma_i$ are the eigenvalues of the $n \times n$ covariance matrix $\mathbf{C}$. (See, e.g., the work of Morgera [32]; two-dimensional case, i.e., mutual information, was derived also by Fraser [14].)

Formula (7), obviously, may be associated with any positive definite covariance matrix. Thus, we use formula (7) to define the *linear redundancy* $L(X_1; \ldots; X_n)$ of an arbitrary $n$-dimensional random variable $X_1, \ldots, X_n$, whose mutual linear dependences are described by the corresponding covariance matrix $\mathbf{C}$:

$$L(X_1; \ldots; X_n) := \frac{1}{2} \sum_{i=1}^{n} \log(c_{ii}) - \frac{1}{2} \sum_{i=1}^{n} \log(\sigma_i). \tag{8}$$

If formula (6) is evaluated using the correlation matrix instead of the covariance matrix, then particularly $c_{ii} = 1$ for every $i$, and we obtain

$$L(X_1; \ldots; X_n) = -\frac{1}{2} \sum_{i=1}^{n} \log(\sigma_i). \tag{9}$$

# 4 Redundancy-based test for nonlinearity: qualitative method

In many experimental situations one records a time series $\{y(t)\}$ of a specific observable. $\{y(t)\}$ is usually considered as a realization of a stochastic process $\{Y(t)\}$ which is stationary and ergodic.

We will study the redundancies for the variables

$$X_i(t) = y(t + (i-1)\tau), \ i = 1, \ldots, n, \tag{10}$$

where $\tau$ is a time delay and $n$ is an embedding dimension [50]. Redundancies of the type
$$R(y(t); y(t+\tau); \ldots; y(t+(n-1)\tau))$$
are, due to stationarity of $\{Y(t)\}$, independent of $t$. We introduce the notation:

$$R^n(\tau) := R(y(t); y(t+\tau); \ldots; y(t+(n-1)\tau)) \tag{11}$$

for the redundancy and

$$L^n(\tau) := L(y(t); y(t+\tau); \ldots; y(t+(n-1)\tau)) \tag{12}$$

for the linear redundancy of the $n$ variables $y(t), y(t+\tau), \ldots, y(t+(n-1)\tau)$. (Quantities (11) and (12) are obtained from a single process realization — an experimental time series — by time averaging, which can be applied due to the above requirement of ergodicity.)

The linear redundancy, according to eq. (9), reflects dependence contained in the correlation matrix **C** of the variables under study. In the special case considered here, when all the variables are, according to (10), lagged versions of $y(t)$, each element of **C** is given by the value of the autocorrelation function of the series $\{y(t)\}$ for a particular lag. As the correlation is a measure of linear dependence, the linear redundancy characterizes only linear structures in the data under study.

Paluš et al. [35] recently proposed to compare the linear redundancy $L^n(\tau)$ with the redundancy $R^n(\tau)$, considered as functions of the time lag $\tau$. If their shapes are the same, or very similar, a linear description of the process under study should be considered sufficient. Large discrepancies suggest important nonlinearities in the dependence among the variables, or, recalling (10), among the studied time series and its lagged versions, i.e., in the dynamics of the process under study.

In this particular approach shapes of redundancies as functions of the lag $\tau$ are compared, not particular values of the redundancies, as redundancy $R^n(\tau)$ and linear redundancy $L^n(\tau)$ have different numerical properties. While $L^n(\tau)$, for particular $n$ and $\tau$, depends on the time series length only, estimated values of $R^n(\tau)$ depend on a numerical procedure used ("quantization" – namely, the finer the partition of the state space, the higher the estimated values of $R^n(\tau)$, see Appendix A), therefore it is impossible to apply any formal statistical procedure to decide whether the redundancies are significantly different in the quantitative sense. The shapes of $\tau$-plots of $R^n(\tau)$, however, are usually consistent for a large extent of numerical parameters used in the redundancy estimations. Therefore comparison of $R^n(\tau)$ vs. $L^n(\tau)$, in the qualitative sense, is possible and can be useful in assessing whether the time series under study is linear or nonlinear. The method is demonstrated below by several numerical examples. (See also [35], [36].)

The linear redundancy $L^n(\tau)$ and the redundancy $R^n(\tau)$ computed from a time series, generated by a linear autoregression (for description of used data see Appendix B), are presented in Figs. 1a and 1b, respectively. It can be seen that there is no qualitative difference between Figs. 1a and 1b, they are practically the same. Consistent with the origin of the data, comparison of $L^n(\tau)$ and $R^n(\tau)$ suggests that all the dependences in the data dynamics are detectable on the linear level.[1]

The same conclusion can be drawn from comparison of Figs. 1c and 1d, depicting $L^n(\tau)$ and $R^n(\tau)$, respectively, computed from a realization of a linear stochastic process, constructed from a given spectrum with random phases ("surrogate data", see Sec. 5).

Figures 2a and 2b present the linear redundancy $L^n(\tau)$ and the redundancy $R^n(\tau)$, respectively, computed from the time series generated by the Lorenz system in a chaotic state [26]. $L^n(\tau)$ decreases quickly to values close to zero and detects no dependence for $\tau > 10$ samples, while $R^n(\tau)$ detects nonlinear dependence, the level of which oscillates with $\tau$ and is characterized by a long-term decreasing trend. These differences can be characterized as qualitative and suggest that a linear description of the data is insufficient.

Figures 2c and 2d illustrate $L^n(\tau)$ and $R^n(\tau)$, respectively, computed from the time series generated by the Rössler system in a chaotic state [43]. While both the redundancies are of similar oscillating nature, the linear redundancy $L^n(\tau)$ is not able to detect the long-term decreasing trend, clearly seen in the redundancy $R^n(\tau)$. This difference we again consider as important or qualitative. (In fact, this long-term decreasing trend is a specific nonlinear property, actually a signature of chaos, as discussed in [35, 36, 13].)

---

[1] We have obtained the same qualitative results for both the Gaussian and non-Gaussian driving noise used in the series generation. Effects of non-Gaussian driving noise, however, can be seen in the quantitative analysis – see Sec. 6. Also, the effects of a non-Gaussian distribution of the scrutinized data and a possibility of their correction are discussed in Sec. 6.

In summary, in the qualitative comparison of $L^n(\tau)$ with $R^n(\tau)$ (or, equivalently, of $R^n(\tau)$ computed from the scrutinized data and $R^n(\tau)$ from its surrogates, defined in the following section) as functions of the time lag $\tau$, the following phenomena we consider as qualitative differences:

- Beyond a lag $\tau$, $L^n(\tau)$ is (approximately) zero, i.e., no linear dependence is detected, while $R^n(\tau)$ is positive and clearly detects a nonlinear dependence among $y(t), y(t+\tau), \ldots, y(t+(n-1)\tau)$. (Cf. Figs. 2a and 2b).

- Both $L^n(\tau)$ and $R^n(\tau)$ are positive in a particular range of $\tau$, however, "dependence structures" reflected in the redundancy vs. lag curves contain clear incompatibilities, like different $\tau$-positions of extrema of $L^n(\tau)$ compared with those of $R^n(\tau)$, or the extrema of $R^n(\tau)$ show a periodic structure while the extrema of $L^n(\tau)$ are located more or less randomly. (Cf. Figs. 10a and 10b).

- There is a long-term decreasing trend detected by $R^n(\tau)$ which is not reflected in $L^n(\tau)$ (cf. Figs. 2c and 2d); or, both $L^n(\tau)$ and $R^n(\tau)$ detect long term trends, but the trends are different (e.g., slower, close to linear decrease in $R^n(\tau)$ and faster, exponential or power-law decrease in $L^n(\tau)$ – see, e.g., Fig. 9 in [36]).

## 5 Quantitative testing: surrogate data

The redundancy-based method, introduced in [35] and reviewed in Sec. 4, was called "the test for nonlinearity". A reader with background in mathematical statistics could argue that "a test" means a precisely elaborated method for testing well formulated null hypotheses by evaluating some statistical quantity, the value of which discriminates whether the null hypothesis should be rejected or not. As was discussed above, the compared quantities — the linear redundancy $L^n(\tau)$ and the redundancy $R^n(\tau)$ have different numerical properties and derivation of some statistical quantity, able to discriminate, whether $L^n(\tau)$ and $R^n(\tau)$ are significantly different, is practically impossible. In order to find a way to a quantitative statistic we must incorporate another idea – the concept of "surrogate data", advocated by Theiler et al. [53].

Surrogate data are artificially generated data which mimic some of the properties of the data under study, but not the property which we are testing for. In the case of testing for nonlinearity, the surrogate data should have the same spectrum and autocorrelation function ("linear properties") as the original data under study, however, surrogate data are generated as realizations of a linear stochastic process. It can be achieved in the following way: Compute the Fourier transform (FT) of the original data, randomize the phases but keep the original absolute values of the Fourier coefficients (i.e., the spectrum) and perform the inverse FT into the time domain. The resulting time series is a realization of a linear stochastic process with the same spectrum as the original data.

Another way to generate a linear stochastic surrogate is fitting an ARMA (autoregressive moving-average) model. Theiler et al. [53] discuss relations between the FT- and ARMA-based surrogates and argue that, for testing hypotheses, the FT-based surrogates are better. Theiler and Prichard [56] demonstrate that a test for nonlinearity based on the FT surrogates can be more powerful than the same test based on the AR surrogates, and an actual nonlinearity in a data can be neglected by a test using the latter. In the following we will focus ourselves on the FT surrogates.

In the case of testing for nonlinearity, the null hypothesis is a linear stochastic process and for a statistic one can evaluate some nonlinear quantity, e.g., correlation dimension/integral or nonlinear forecastibility for the original data and a set of the surrogate data (different realizations of the linear stochastic process, i.e., a set of series obtained by the inverse FT of the same spectrum but different sets of random phases). The resulting statistic then is the difference between the value obtained from the scrutinized data and the mean value obtained from the set of surrogates, divided by the standard deviation (SD) of the surrogates, i.e., the difference in number of SD's. The null hypothesis is rejected and the result called *significant*, if the probability $p$ of the null hypothesis is lower than a chosen level, usually set to 0.05 or 0.01. Providing the statistic has a normal distribution $N(0,1)$ and 30 realizations of the surrogates were used in the test[2], then probability of the null hypothesis is $p < 0.05$ for values of the statistic greater than 1.699; or greater than 2.462 for $p < 0.01$, etc., see, e.g., [3].

The above critical values of a statistic are related to one-sided tests, because the change of the statistic from a possible nonlinear deterministic process to a linear stochastic process is expected in one direction (e.g., increase of dimensionality). As we will see later, this is not always true, however, the anomalous changes of a statistic are usually caused by numerical artifacts and should be handled by a different way than a small increase of the critical values according to two-sided tests.

---

[2] With a limited number $N_S$ of the realizations of the surrogates, the t-distribution with $(N_S - 1)$ degrees of freedom should be used for deriving the critical values of statistics, instead of the normal distribution. In all the examples, presented here, 30 realizations of surrogates were used, therefore the critical values presented in this paper are related to the t-distribution with 29 degrees of freedom. These are, however, close to those related to the normal distribution, e.g., for $p = 0.05$ the critical value is 1.699 for the former and 1.645 for the latter.[3]

If the difference is significant, one should reject the null hypothesis (a linear stochastic process) and consider the result as a signature of nonlinearity or nonlinear determinism in the data under study. (For more details and examples see Theiler et al. [53].)

Let us outline what are the relations among the redundancies $R^n(\tau)[o]$ and $L^n(\tau)[o]$, computed from the original data and the redundancies $R^n(\tau)[s]$ and $L^n(\tau)[s]$, computed from the surrogate data.

The surrogate data are linear by the way of their construction, which implies the equality $L^n(\tau)[s] = R^n(\tau)[s]$ (we mean equality in the above considered sense, i.e., the shapes of the redundancies as the functions of $\tau$ are the same). The surrogate data have the same spectrum as the original one, that means they should have also the same autocorrelation function[3] (for more details see [55]). Recalling the relation between the autocorrelation function and the linear redundancy, we can see that: $L^n(\tau)[s] = L^n(\tau)[o]$. This means that $R^n(\tau)[o] = L^n(\tau)[o]$ if and only if $R^n(\tau)[o] = R^n(\tau)[s]$. That is, instead of comparing $R^n(\tau)[o]$ with $L^n(\tau)[o]$ we can compare $R^n(\tau)[o]$ with $R^n(\tau)[s]$. Thus the compared quantities have the same numerical properties and, provided we use the same numerical parameters in the estimations of $R^n(\tau)[o]$ and $R^n(\tau)[s]$, their values can be compared quantitatively.

Alternatively, we could simply propose the redundancy $R^n(\tau)$ as a nonlinear quantity suitable for deriving a discriminating statistic in the quantitative evaluation based on the surrogate data method. However, the above introduced concept of the linear redundancy $L^n(\tau)$ and the qualitative comparison still can be useful as far as the quantitative testing is not always reliable, as we will demonstrate below.

It is well-known, that in construction of $n$-dimensional embeddings by the time delay method, the results are usually influenced by the choice of the embedding dimension $n$ and the time delay $\tau$. Therefore, also in the quantitative testing, we evaluate the redundancy-based statistics for several embedding dimensions, usually $n = 2, \ldots, 5$ and a broad range of time delays.

(This approach, yielding multiplicity of test values, opens a question of simultaneous statistical inference, which we will discuss in connection with a particular practical application of the redundancy-based nonlinearity testing (Sec. 8, [37]), rather than here. For more information see, e.g., [33, 29, 31, 19].)

---

[3]In some cases it is not true, due to numerical reasons, see Sec. 6.

The usefulness of plotting a statistic versus time lag to illustrate differences between test and surrogate data was also demonstrated by Kaplan and Glass [22].

# 6  Spurious nonlinearity - pitfalls in the surrogate data

Theiler et al. [55] recently have observed that generating a surrogate data from a time series with a long coherence time, i.e., from strongly autocorrelated series, can be problematic, and formal application of a statistical test can lead to false detection of nonlinearity. In the following illustrative example we demonstrate how this problem can be attacked by the redundancy-based analysis.

Figure 3 presents the results of the quantitative testing of the data, qualitative comparison of which was presented in Figs. 1c, 1d. The data is an example of a linear stochastic process by construction, in this case a particular realization of the surrogate of the series generated by the chaotic Rössler system. The time series length used for the data generation and then for the subsequent analysis, presented in Figs. 3a and 3b, was 131,072 (128K) samples. According to expectation, there is no significant difference between the original data and its surrogates, differences oscillate between -1 and 1 SD as measured by both $L^n(\tau)$ and $R^n(\tau)$, respectively. (Of course, the absolute values of these differences are very small, of the order of $10^{-3}$ in redundancies and $10^{-4}$ in related autocorrelation functions (ACF) for both the unbiased and circular estimators [20, 55].)

The situation is different, when we analyze only 16,384 (16K) samples from the above data, i.e., now we construct 16K-surrogates from a (16K) "test data", which is, in fact, a 16K segment of a surrogate generated by FFT of 128K samples. In Fig. 3d one can see that differences in redundancies $R^n(\tau)$ reach values over 1.7 SD's, for small $\tau$ even more than 2 SD's, so one could formally reject the null hypothesis of the linear stochastic origin of this data. However, looking at differences in Fig. 3c, derived from the linear redundancies $L^n(\tau)$, one can see that 16K-sample surrogates are significantly different from the scrutinized data on the linear level, too. This means that the surrogate data, in spite of theoretical expectation, does not exactly mimic the linear properties of the original series, and this is the reason of significant differences indicated by the nonlinear statistic, not a nonlinearity in the data.

The linear redundancy-based statistic, presented in Fig. 3c, was computed using the unbiased estimator of the ACF. When the circular estimator was used, the differences are slightly smaller, but still significant. Also, behavior of other 16K segments of the full 128K "test data" was very similar.

Deriving the statistic from the ACF itself (by the same way as from the redun-dancies) we found consistent decrease of (absolute) values of the ACF, however, the decrease is not significant (maximum difference is about 1.3 SD). Thus, the linear redundancy $L^n(\tau)$, as a transformation of the ACF, is more sensitive to the differences in the surrogates than the ACF itself. As we have demonstrated above, $L^n(\tau)$ is equivalently sensitive as the (nonlinear) redundancy $R^n(\tau)$, which is the important feature of this technique — consider a nonlinear statistic, at least as sensitive as the statistic based on $R^n(\tau)$, used in a nonlinearity test and a possibility of a spurious nonlinearity detection caused by imperfect surrogates is tested by the ACF. Then a researcher, relying on significant results only (i.e., neglecting the decrease in the ACF because it is not significant), can evaluate the above linear stochastic series as nonlinear.

Considering the question why the linear properties of the test data are not preserved in the surrogates, Theiler et al. [55] argue, that one of the possible reasons can be that the true power spectrum of coherent signals, like that presented in this example, contains sharp spikes, which, when estimating the power spectrum from a finite time series, are spread out over only several frequency bins and thus loose their "sharpness". As a consequence, the surrogate data have shorter coherence time and weaker autocorrelation, which leads to lower values of redundancies. If this is the correct explanation, then, in the above example, the "sharpness" of the dominant peak was changed by using shorter time series than originally used in generation of the data under study.

(Also, surrogate data of a periodic series can be distorted when a time series length is not commensurate with the period(s) of the series, see Theiler et al. [55].)

Unfortunately, the sharp-peak spectrum is not the only source of pitfalls in the surrogate data. Even in the quantitative analysis of the AR process (Gaussian driving noise) from Figs. 1a, 1b we have observed significant (1.7 - 1.9 SD's) decrease of redundancies of its surrogates in comparison with redundancies of the original data. (However, this series has the long coherence time, too.)

We have also observed the opposite effect: Fig. 4 presents the redundancies computed from a noisy sine signal and its surrogates. In this case one can see that the surrogates give higher values of both the redundancies, i.e., they have stronger autocorrelations that the original series. (The differences are in order of tens of SD's.) This effect usually emerges in series with a large amount of high-frequency noise, i.e., with the signal-to-noise ratio equal to or less than 1 (the sine with a small portion of noise can give different results – see Sec. 7), and when a noise has high amplitudes also in the high-frequency cut-off of a spectrum, i.e., a low-pass filtering, or, in numerically generated data, using a 1/f noise can prevent this effect.

Increase of redundancies (or, equivalently, possible decrease of a dimensional or forecasting-error statistic) in the surrogates is clearly an anomalous result and one can be aware that there is something wrong with the surrogates, unless he/she records only the absolute values of a statistic. Then also this phenomenon can lead to a spurious nonlinearity detection.

Surrogate data with linear properties different from the original data, yielding either decrease or increase of redundancies, have been observed extensively also in processing of experimental data, especially human electroencephalographic signals (to be published).

We have no ambition to find an explanation (or explanations) why these flaws in the surrogates can occur. The aim of this paper is to warn researchers that the generation of the FT-based surrogate data is not always a process preserving linear properties of an original data. Changes of linear properties, like the decrease of autocorrelations, are reflected in nonlinear measures as well and thus can lead to false detection of nonlinearity. In order to avoid such spurious detection of nonlinearity, one should test differences between the data and its surrogates on both the linear and nonlinear levels, using adequate tools. And the redundancy twins $L^n(\tau)$ and $R^n(\tau)$ could be suitable for such a double testing approach. When in a particular study only the redundancy $R^n(\tau)$ yields significant differences between the scrutinized data and its surrogates and there is no significant difference in the statistic based on the linear redundancy $L^n(\tau)$, a possibility of spurious effects, due to change of the surrogates' linear properties, can be rejected. However, when also the linear redundancy $L^n(\tau)$ yields significant differences, then the nonlinear character of the data under study cannot be taken for granted.

There is, however, another problem: Nonlinear and even chaotic series can have long coherence times and strong linear autocorrelations, which can be decreased in the surrogate data, i.e., using the above "double testing", actual nonlinearity can be considered doubtful. This is the case of the chaotic series generated by the Rössler system (Figs. 2c, 2d). The quantitative analysis of the time series generated by the chaotic Rössler system brought highly significant differences on both the linear (the statistic based on $L^n(\tau)$) and nonlinear (the statistic based on $R^n(\tau)$) levels. Thus in this case the only safe way of detection of the nonlinear character of this time series is the qualitative comparison of $L^n(\tau)$ and $R^n(\tau)$, as it was presented in Sec. 4.

Another possible source of spurious detection of nonlinearity in a dynamics of a scrutinized time series can be a *static* nonlinearity: Let us suppose that the underlying dynamics of the studied system is linear and Gaussian, i.e., there is an original realization $\{x_i\}$ of a Gaussian process, but we can measure the series $\{y_i\}$, $y_i = f(x_i)$, where $f$ is a monotonic nonlinear function. Such nonlinearity, which is not intrinsic to the dynamics of the system under study, but can be caused, e.g., by a measurement apparatus, we call static nonlinearity. The realization of the linear stochastic process (a surrogate of the Rössler chaotic series), studied above (Figs. 1c, 1d – the qualitative analysis, Figs. 3a, 3b – the quantitative testing) was passed through the static exponential nonlinearity. The quantitative analysis of this series is presented in Figs. 5a and 5b. While there is no significant difference between the data and its surrogate on the linear level, i.e., measured by the linear redundancy $L^n(\tau)$ (Fig. 5a), the nonlinear statistic based on the redundancy $R^n(\tau)$ yields significant results – up to 4 SD's (Fig. 5b). Thus, one should formally reject the hypothesis of a linear stochastic origin of the process. The static nonlinearity can also distort the results of the qualitative analysis, as presented in Figs. 5c, 5d. (This is, however, not a rule, i.e., not all static nonlinearities distort the qualitative analysis, as was shown in [36].)

A static nonlinear transformation of a time series, or, a static nonlinearity, as it was called above, effectively means that the distribution of the data is not Gaussian (even if the underlying process is Gaussian). So let us perform another nonlinear transformation of the data - "gaussianization", or "histogram transformation" in order to obtain a Gaussian distribution of the scrutinized data. Due to the central-limit theorem, the probability distribution of the FT-based surrogate data tends to a Gaussian distribution. Therefore, this can be the one way to eliminate influence of the static nonlinearity on the nonlinearity test.

For this purpose, we create a sample of Gaussian random numbers of the same length as the scrutinized series, then sort both the data sets in numerical order, making a one-to-one correspondence between points with the same index. This defines an invertible nonlinear transformation of the original time series to a new one, with an approximate Gaussian distribution.

This transformation simply removes the effect of the static nonlinearity. Analysis of the gaussianized time series, containing the static nonlinearity, yields the same results as the original linear series, i.e., those presented in Figs. 1c, 1d, 3a and 3b.

Performing the "gaussianization" of the analyzed data could be a useful step in order to prevent effects of static nonlinearities, or non-Gaussian distributions of the data under study, on the nonlinearity tests and thus to detect only true dynamical nonlinearity. (This problem is also discussed by Kennel and Isabelle [23], Theiler et al. [53] call an equivalent procedure "amplitude adjustment", Rapp et al. [42] use the term "Gaussian-scaled surrogates".)

An interesting question emerged when we consider the above AR process con-structed with a non-Gaussian driving noise. As was mentioned in Sec. 4 even with the non-Gaussian driving noise the qualitative comparison of $L^n(\tau)$ with $R^n(\tau)$ yields no qualitative differences. In the quantitative testing, however, the situation is different. Let us remark first, that even with the Gaussian driving noise we obtained slightly significant (1.7 – 1.9 SD's) differences between the data and the surrogates, detected also on the linear level, which we characterized above as a flaw in the surrogates. On the other hand, when as the driving noise we used the Gaussian noise passed through a static nonlinear (quadratic or exponential) transformation, we have observed increase in the nonlinear (redundancy) statistic up to 6 SD's. (This increase was not observed in the linear statistic.) Moreover, this significant difference was not attenuated in the gaussianized data, so we should admit a dynamical nonlinearity in the series. In fact, it is true – although the driving noise itself underwent just the static nonlinear transformation – this nonlinearity "entered" the dynamics of the series, as far as each element of the series depends on its several predecessors. In any case, such a process is compatible neither with the null hypothesis of a linear stochastic process (see Sec. 2), nor with the hypothesis of a linear stochastic process transformed by a static nonlinearity. Recalling that we have found no differences in the qualitative comparison, we can suspect that this nonlinearity is not a very "dynamically interesting" phenomenon. (See also the following discussion on the "trivial" nonlinearity.)

# 7 Qualitative versus quantitative testing: a trivial nonlinearity

Above we have discussed possible sources of spurious identification of nonlinearity in time series. Although we were concentrated on the redundancy technique, introduced in Sec. 3, 4 and 5, these phenomena can influence any nonlinearity test. (See also [53, 55].) The redundancy technique, however, offers unique tools for testing both the linear and nonlinear properties of the scrutinized data and its surrogates and also for their quantitative and qualitative comparisons.

Let us consider the situation when the quantitative analysis yields the "safe" detection of nonlinearity (i.e., free from spurious results due to linear distortions in the surrogate data), however, there is no difference in the qualitative comparison. The former implies rejection of the null hypothesis of a linear stochastic process, while the latter suggests, that all the dependence structures in the scrutinized data are mimicked qualitatively by a linear stochastic process.

Let us consider the linear system

$$dx/dt = \omega y,$$

$$dy/dt = -\omega x,$$

a solution of which is $x = \sin(\omega t)$, $y = \cos(\omega t)$. Let us record the $x$-component of the system, i.e., the sine wave, with a moderate amount (10%) of additive Gaussian noise. (In this and the following example we use an additive noise in order to avoid effects of aliasing of sampling frequency, resulting in spurious peaks in redundancy graphs.) This signal is the output of a linear system and, in principle, can be considered as a realization of a linear stochastic process. On the other hand, it is deterministic and the relation between $\sin(\omega t)$ and $\sin(\omega(t + \tau))$ is nonlinear.

The results of the analysis of this noisy sine series 128K samples long are presented in Fig. 6. The linear redundancy statistic (Fig. 6a) gives values of about 0.5 SD's, i.e., no significant difference from the surrogates on the linear level, while the (nonlinear) redundancy statistic (Fig. 6b) yields the values up to 5 – 6 SD's - i.e., significant differences from a linear stochastic process.

These results support the proposition that the signal is nonlinear and deterministic. However, one can defend the assertion that the signal is a realization of a linear and stochastic process, that can be constructed from the spectrum/periodogram in which only one frequency bin is nonzero, or, more exactly, only one frequency bin has a distinctly higher value than a white-noise background.

The periodogram of the noisy sine series is presented in Fig. 7. We can see that there is a single peak over a white-noise background, however, more detailed look shows that the peak is spread out over several hundreds (from 65,536) of frequency bins.[4] A particular nonlinear determinism of the series is encoded in the phase relations among these frequency bins. In the surrogate data the phases are randomized and thus we get the results from Fig. 6b. If there was only one bin over the noise level, a particular set of phases would not play any important role. Therefore one can insist on linearity of the series and our result qualify as a numerical artifact, i.e., the nonlinearity detected can be considered as spurious.

Such a discrete-spectrum linear stochastic process, obviously, is not typical, or realistic. Therefore we consider the explanation of the series by the nonlinear and deterministic process as the correct one, and consider our results not spurious, i.e. nonlinearity detected in this noisy sine series we consider as a true dynamic nonlinearity. On the other hand – Fig. 6d presents $R^2(\tau)$ of the data and its surrogates (let us remind that the qualitative comparison of $L^n(\tau)[o]$ with $R^n(\tau)[o]$ is equivalent to that of $R^n(\tau)[o]$ with $R^n(\tau)[s]$) – we can see that the only differences are quantitative while the qualitative picture of the dependences in the series are mimicked well by the linear stochastic surrogates. In other words, this nonlinearity does not induce any specific nonlinear ("dynamically interesting") behavior. Therefore, for this kind of nonlinearity we propose the term *trivial nonlinearity*. In this case the series was even generated by a linear system. Detection of trivial nonlinearity, however, does not mean that the system under study is linear. In Fig. 8 we present results of analysis of the series generated by the Rössler system in a periodic state [30] (again recorded with a small amount of the additive Gaussian noise). The results, again obtained from the series 128K samples long, are the same as those for the noisy sine series: no significant differences on the linear level (Fig. 8a) – the quantitative analysis, Fig. 8c – the qualitative analysis – $L^2(\tau)$ from the data and the surrogates coincide), highly significant quantitative differences in the (nonlinear) redundancy statistic (Fig. 8b), however, no qualitative differences between $R^n(\tau)$ from the data and the surrogates (Fig. 8d).

---

[4]This is the effect of finite precision, i.e. in numerical practice even using deterministic sine function we are not able to generate a series with exactly discrete spectrum.

The series analyzed above have obviously long (in fact, infinite) coherence lengths. Isn't the "trivial" nonlinearity in fact the spurious one, due to a long coherence length, as demonstrated in Sec. 6? Let us investigate in detail the linear stochastic process analyzed in Sec. 6, Fig. 3, i.e., the 128K-sample surrogate of the Rössler chaotic series. Here we analyze the series subset, 16K samples long, as in Figs. 3c and 3d. The linear redundancies $L^2(\tau)$ and the redundancies $R^2(\tau)$ computed from the data and the surrogates are presented in Figs. 9a and 9b, respectively. We can see that the largest differences between the value of the redundancy of the original data and the mean value of the redundancy of the surrogates are located in the maxima of both $L^2(\tau)$ and $R^2(\tau)$. On the other hand, in the cases of the above analyzed periodic data (the sine and the periodic Rössler series) these differences are largest in the minima of $R^2(\tau)$ (Figs. 6d, 8d), the curves of $L^2(\tau)$ from the data and surrogates are almost identical.

Figures 9c and 9d present $L^2(\tau)$ and $R^2(\tau)$, respectively, computed from the above periodic Rössler series and its surrogates, like in Figs. 8c and 8d, in this case, however, from the subseries 16K samples long. The spurious decrease of redundancies/autocorrelations, discussed in Sec. 6, emerged here: $L^2(\tau)$ of the data and the surrogates (Fig. 9c) are not more indistinguishable, like in Fig. 8c, but differences in maxima of $L^2(\tau)$ appeared, like in Fig. 9a, i.e. the change in linear properties emerged. It means that by using a long enough series (128K samples) the above spurious effect (i.e., change in linear properties) was avoided and the nonlinearity detected in Figs. 6b and 8b is not spurious. Redundancy $R^2(\tau)$ in Fig. 9d detects both the changes - in the linear properties (maxima) and in the nonlinear properties (minima) as in the case of series 128K samples long.

This analysis gives us more detailed insight into the character of the above described trivial nonlinearity. The functions $\sin(\omega t)$ and $\sin(\omega t + \pi/2)$ are linearly independent (uncorrelated). On nonlinear level, however, there is a clear deterministic relation. So $L^2(\tau)$ for the above periodic series (the noisy sine and the noisy periodic Rössler series) and the lag $\tau = \pi/2$ (and its odd multiples) has the zero value, while $R^2(\tau)$ is clearly above zero indicating nonlinear dependence. In the surrogates only linear dependences are preserved - i.e., both the $L^2(\tau)$ and $R^2(\tau)$ are the same for the data and the surrogates in or close to the maxima (i.e., the lags $\tau = \pi, 2\pi, \ldots$), while in the minima $R^2(\tau)$ is zero for the surrogates and nonzero for the data. This is the consequence of the nonlinear deterministic relation in the data (i.e., the particular phase relations), destroyed in the surrogates. (Also, the largest differences in Figs. 6b and 8b were obtained from $R^2(\tau)$, i.e., for the embedding dimension $n = 2$, when the relation of two (linearly independent) variables ($\sin(\omega t)$ and $\sin(\omega t + \pi/2)$) is evaluated; for $n = 3$ (and higher) the differences are smaller, because another variable, shifted by $\pi$ relatively to the first variable, enters the analysis and thus there is also some level of linear dependence among the variables.)

The reason why we call this phenomenon *trivial* nonlinearity is an effort to distinguish this type of dynamics[5] from the specific nonlinear behavior like an evolution on chaotic and/or fractal attractors. As we demonstrated in Fig. 2, such a dynamics evokes qualitative differences in the $L^n(\tau) - R^n(\tau)$ comparison. Also, in the quantitative testing the differences are significant for a whole interval of time lags, not only for particular lag values (the minima of $R^2(\tau)$) as it is in the case of the trivial nonlinearity. In Fig. 10 we present results of the analysis of another example of "nontrivial" nonlinearity. The system, introduced by Grebogi et al. [18], is an example of the evolution on a "strange nonchaotic attractor", i.e., the attractor is a fractal set, but the (3-dimensional) system has one zero and two negative Lyapunov exponents. (The distinction chaotic – nonchaotic, as it is reflected in redundancies, is discussed in [36].) The redundancy $R^n(\tau)$ (Fig. 10b) reflects a clear periodic structure, which is not detectable on the linear level, i.e., in $L^n(\tau)$ (Fig. 10a). In the quantitative testing, there is no significant difference on the linear level, i.e., as measured by $L^n(\tau)$ (not presented), while in the nonlinear statistic, based on $R^n(\tau)$, the differences are from 5 to 20 SD's, i.e., significant for any lag $\tau$ (Fig. 10c). Fig. 10d presents $R^2(\tau)$ from the original data and the mean value from the surrogates. We can see clear distinctions in both quantitative and qualitative ways.

---

[5] Or types of dynamics, as we stated in Sec. 6, a different example of a trivial nonlinearity can be an AR process driven by a non-Gaussian noise. We have focused our attention to the above kind of trivial nonlinearity, resulted from periodic dynamics, just because we have observed it extensively in experimental data, some of which were previously characterized as chaotic.

Let us suppose, in the following, that the data under study have a Gaussian distribution, or, was "gaussianized", as proposed above, so that the possibility of spurious effects of a static nonlinearity was eliminated. Table 1 summarizes possible outcomes of the quantitative tests on both the linear and nonlinear levels, combined with the results of the qualitative comparison. Interpretations for all eight combinations are given.

===== Table 1. — approximate location =====

Let us explain interpretations of the results in details.

1. If the results obtained are not significant in both the linear (based on $L^n(\tau)$) and nonlinear (based on $R^n(\tau)$) statistics and no differences were found in the qualitative comparison, then we characterize the data under scrutiny as linear. To be more precise, we didn't prove that the data is linear, we use the term "linear" for more exact "indistinguishable from a linear stochastic process".

2. Non-significant results in the quantitative testing, but qualitative differences — we have never observed this combination of results. Although we do not have a theoretical evidence, based on extensive numerical experience, we can conjecture, that qualitative differences imply significant quantitative differences.

3. Results are significant in the linear statistic, but non-significant in the nonlinear statistic: We list this option only formally; it is not real, since the redundancy $R^n(\tau)$ is sensitive also to (changes in) linear dependence.

4. The same conclusion as in the previous item.

5. If the quantitative results are not significant in the linear statistic and significant in the nonlinear statistic, we have quantitative detection of nonlinearity safe from spurious effects of imperfect surrogates. In combination with NO qualitative differences, however, we suspect that this nonlinearity is not "dynamically interesting", or very substantial for the studied dynamics. We call this case a "trivial nonlinearity" in the sense discussed above.

6. Results are not significant in linear, but significant in nonlinear statistic and also qualitative differences were found: The data was generated by a specifically nonlinear dynamics such as an evolution on chaotic and/or fractal attractors.[6]

---

[6]This is usually the most interesting case of nontrivial nonlinear dynamics, but probably not the only case. Evidence for nontrivial nonlinearity does not mean automatically evidence for chaos.

7. Quantitative results significant in both the linear and nonlinear statistics and no qualitative difference: In this case the data can be either linear (significances are caused by flaws in surrogates), or nonlinear (in the surrogates, nonlinear dependence was destroyed, but also linear properties are not mimicked entirely). Due to no qualitative differences, however, the possible nonlinearity is the "trivial" one.

8. Quantitative results significant in both the linear and nonlinear statistics and qualitative differences found: This is the case when the qualitative analysis can solve the problem of unclear result of the quantitative tests and clearly detect specifically nonlinear dynamics, as it was mentioned in the case of the chaotic Rössler system above.

# 8 Examples of real data analysis

In order to demonstrate the applicability of the presented method on short experimental time series, we present results of analysis of two experimental datasets, which were previously analyzed by other authors using different methods. In both the cases we perform gaussianization, as described in Sec. 6, in order to eliminate possible effects of static nonlinearities and thus to detect only dynamical nonlinearity.

*Sunspots*

The annual Wolf sunspot numbers have been recorded since 1700. The time series is described and displayed, e.g., in [57]. We have analyzed the sunspot numbers for the years 1724 – 1979, giving 256 observations. Using such short time series we restricted the test for embedding dimension $n = 2$ only, and lags 1 – 32 years. Redundancies $R^n(\tau)$ were computed using four equiquantal marginal levels (see Appendix A). The results are presented in Fig. 11.

Before stating whether the results are significant or not, we must take into account the problem of simultaneous statistical inference — since we have obtained multiple test results and only some of them seem to have values large enough to be significant, the critical value for the statistic must be adjusted in order to avoid randomly significant results. The simplest way is the following: Having $m$ test values, the whole test result, based on an individual test value, is significant on $p = 0.05$ if the individual test value is significant on $p = 0.05/m$ [29, 33]. If we can expect $k$ significant values (of $m$ test values), 5% significance is given by their significance on $p = 0.05k/m$ [19, 31]. Then, for rejecting the null hypothesis of a linear stochastic process on $p = 0.05$, with $m = 32$ and $k = 1$ the critical value of 1.699 should be increased to the value about 3.4, if we can expect $k = 5$, we can consider the critical value about 2.6.[7]

In the linear (based on $L^2(\tau)$) statistic (Fig. 11a) there are no significant differences detected, while the result of the test performed using the nonlinear (based on $R^2(\tau)$) statistic (Fig. 11b) is significant. Non-significant result in the linear statistic warrant that the significance in the nonlinear statistic is not spurious, so we can reject the null hypothesis and consider the sunspot series to be nonlinear.

For the qualitative test of the sunspot series we present $R^2(\tau)$ of the surrogates (Fig. 11c) and $R^2(\tau)$ of the data (Fig. 11d), rather than $L^n(\tau)$ and $R^n(\tau)$ of the data, since for small amount of data we consider the former more reliable than the latter. Also, the values of these estimates are not very reliable. As we emphasized earlier, in the qualitative comparison we are not interested in the values of the redundancies, however, there is one value - the value of zero redundancy, which can be important even in the qualitative description of the studied dynamics. In order to find a "numerical zero" in this test, we generated 256 Gaussian random numbers and computed the redundancies $R^2(\tau)$ using the same numerical parameters as those used for the sunspot series and its surrogates. The mean value of $R^2(\tau)$ for this random series, and the mean $\pm$ SD are illustrated by dash-and-dotted and dashed lines, respectively, in Figs. 11c and 11d. Thus the values between the two dashed lines represent an "adjusted zero" for the redundancy estimates and detect independence between the series and its lagged twin.

After the adjustment of the zero we can see that $R^2(\tau)$ of the sunspots is nonzero for the whole range of lags, studied here, while $R^2(\tau)$ of the surrogates reaches zero in its minima. This is the behavior we have observed for numerically generated nonlinear series and their surrogates in previous sections.

The redundancies $R^2(\tau)$ of the data and the surrogates, as the functions of the lag $\tau$, are not exactly the same, however, the positions of extrema are consistent. Also, looking at the nonlinear statistic (Fig. 11b) we can see that the significant values are located only in the minima of $R^2(\tau)$. Thus we suspect that detected nonlinearity is closer to a kind of a trivial nonlinearity[8] than to a specific nonlinear behavior like evolution on strange attractors. Therefore we can agree with Suba Rao's conclusion that the sunspot data is nonlinear, but not chaotic. [48]

---

[7]We did not use this adjustment in the tests depicted in Figs. 3c and 3d, since the ratio $k/m$ was close to one. In other tests the test values, considered as significant, were large enough to fulfill the conditions of this adjustment, like values about $5 - 6$ in Fig. 6b, or even 30 in Fig. 8b.

[8]Of course, not exactly the same as the trivial examples presented in Sec. 7. More complicated behavior can be observed, e.g., if similar "trivial nonlinearity" data is generated with frequency which is not constant, but fluctuating.

*Measles*

The number of measles cases reported each month in New York City in the years 1928 – 1963 was studied by Sugihara and May [49] and more recently by Kaplan and Glass [22]. We have analyzed last 256 samples using the same numerical conditions as in the case of sunspots, but for lags 1 – 60 months. The results are presented in Fig. 12. The critical value of the statistic (adjusted to multiple test values) in this case is approximately 2.76.

In the nonlinear (based on $R^2(\tau)$) statistic (Fig. 12b) there are several significant values, however only one of them (lag = 3 moths) is significant only in the nonlinear statistic, while other significances are detected by the linear (based on $L^2(\tau)$) statistic (Fig. 12a) as well. This can indicate that the significances are caused by imperfect surrogates and not by nonlinearity. Before we make any conclusion based on the quantitative test, let us consider also the result of the qualitative comparison of $R^2(\tau)$ of the data (Fig. 12d) and its surrogates (Fig. 12c). At the first sight we can see that there are no qualitative differences. Moreover, $R^2(\tau)$ of the measles data in its minima declines to zero value. (The numerical zero was adjusted by the same way as in the sunspots case.) The first zero redundancy occurs at the lag 9 months. This means that possible nonlinear deterministic relations in the series is not spread further than one year. Thus, coming back to the quantitative test, all the significant results for lags greater than 1 year are probably spurious, i.e., caused by the surrogates which do not mimic exactly linear properties of the original data, what is confirmed by the significant differences, for the same lags, detected by the linear statistic. The only difference, significant in the nonlinear statistic, but nonsignificant in the linear statistic, is that located in the lag 3 months, what is also the first minimum of $R^2(\tau)$.

Considering both the quantitative and qualitative results, we can conclude that for lags not larger than 1 year, in the measles data, there is a nonlinear deterministic relation of the type of trivial nonlinearity (in the sense discussed in Sec. 7), while for all the dependences in the lags larger than 1 year we were not able to reject reliably a linear stochastic explanation.

Sugihara and May [49] analyzed this data by assessing its predictability using linear and nonlinear models, and concluded that it shows signs of deterministic chaos. This conclusion was based on the fact that the prediction accuracy decreased with increasing prediction time. This feature we can consider as necessary, but not sufficient condition for chaos. This behavior was observed in range of 1 – 12 months, with the loss of predictability for longer times, what means, in particular, that one year's level of measles is not predictable from the previous year's. This is consistent with our findings and also with the results of Kaplan and Glass [22], who proposed as a possible model for the measles time series a series of yearly peaks of random amplitudes.

# 9 Conclusion

We have presented the method for testing nonlinearity in time series, based on information-theoretic functionals – redundancies. The *qualitative* comparison, introduced in [35] and in Sec. 4, can neglect some kinds of nonlinearity. Therefore, we introduced also *quantitative* testing as the combination of the redundancy technique and the surrogate data approach [53]. If the question asked is "Is there any nonlinearity or nonlinear determinism in the data?", the quantitative testing is superior, providing that the spurious effects of possible numerical artifacts are avoided. The latter can be achieved by the application of both the linear and general (nonlinear) redundancies for the evaluation of the statistics in the quantitative testing, combined also with the qualitative comparison.

On the other hand, we have shown an example of the nonlinearity, being neglected in the qualitative comparison. This is a kind of the *"trivial"* nonlinearity, like that emerged in simple dynamics on limit cycles or tori, while "dynamically interesting" dynamics, like an evolution on chaotic and/or fractal attractors, can be easily detected by the qualitative comparison, introduced in Sec. 4. The most reliable approach, however, is the combination of both the quantitative and qualitative analyses, using the equivalently sensitive comparisons on both the linear and nonlinear levels, possible due to above introduced redundancy — linear redundancy formalism.

# Acknowledgements

The author would like to thank J. Theiler, D. Prichard, D. Kaplan, D. Chialvo, B. Hinrichs, J. Klaschka and V. Witkovský for valuable comments and inspiring discussions. J. Theiler is acknowledged also for supplying and help with his plotting software.

The author is grateful to D. Kaplan for providing him the measles data and to A. Weigend, who made the sunspots data available on the Santa Fe Institute Time-Series server.

The author is supported by the International Research Fellowship F05 TW04757 ICP from the National Institutes of Health, the Fogarty International Center, and also by grants to the Santa Fe Institute, including core funding from the John D. and Catherine T. MacArthur Foundation, the National Science Foundation (PHY-8714918), and the U.S. Department of Energy (ER-FG05-88ER25054).

## Appendix A

The linear redundancies were computed according to eq. (9). Eigenvalues of each correlation matrix were obtained by using the SVDCMP routine described in [38], p.52. The algorithm for computing the redundancy, proposed by Fraser [13, 14] or Fraser and Swinney [12], is rather complicated one. We have found a simple box-counting method is sufficient. The only "special prescriptions", based on our extensive numerical experience, concern the way of data quantization:
a) The type of quantization: We propose to use the marginal equiquantization method, i.e., the boxes for box-counting are defined not equidistantly but so that there is approximately the same number of samples in each marginal box.
b) The number of quantization levels (marginal boxes): We have found that the requirement for the effective[9] series length $N$ using $Q$ quantization levels in the computation of the $n$-dimensional redundancy is

$$N \geq Q^{n+1},$$

otherwise the results can be heavily biased. Usually better results are obtained with $Q < N^{1/(n+1)}$ than with $Q > N^{1/(n+1)}$. Redundancies computed with $Q < N^{1/(n+1)}$ can be underestimated, but graphs of $R^n$ vs. $\tau$ are, even for $Q = 4 - 6$, similar to those obtained from long time series and $Q = N^{1/(n+1)}$. In the case of $Q > N^{1/(n+1)}$ the redundancies can be overestimated and the curves $R^n(\tau)$ can be distorted.

The numbers $Q$ of the equiquantal marginal levels, used in the computations yielding the presented results are:
$Q = 12$ for the results presented in Figs. 1d and 2d,
$Q = 8$ for the results presented in Figs. 2b, 3b,d, 5b,d, 10b,c,d,
and $Q = 4$ for the rest of the presented results of the redundancy $R^n(\tau)$ and $R^n(\tau)$-based statistic.

---

[9]Having $N_{total}$ data samples, for maximum embedding dimension $n_{max}$ and maximum lag $\tau_{max}$, the effective series length is $N = N_{total} - (n_{max} - 1)\tau_{max}$.

# Appendix B

The realization of the AR process was generated according to the equation:

$$x_t = \sum_{k=1}^{10} a_k x_{t-k} + \sigma e_t,$$

where $a_{k=1,..,10} = 0, 0, 0, 0, 0, .19, .2, .2, .2, .2$, $\sigma = 0.01$ and $e_t$ are Gaussian deviates with zero mean and unit variance.

The data from continuous nonlinear dynamical systems were generated by numerical integration based on the Bulirsch-Stoer method [38] of the chaotic Lorenz system [26]:

$$(dx/dt, dy/dt, dz/dt) = (10(y-x), 28x - y - xz, xy - 8z/3),$$

with initial values (15.34, 13.68, 37.91), integration step 0.04 and accuracy 0.0001; the chaotic Rössler system [43]:

$$(dx/dt, dy/dt, dz/dt) = (-z - y, x + 0.15y, 0.2 + z(x - 10)),$$

with initial values (11.120979, 17.496796, 51.023544), integration step 0.314 and accuracy 0.0001; and the periodic Rössler system [30]:

$$(dx/dt, dy/dt, dz/dt) = (-z - y, x + 0.2y, 0.2 + z(x - 2.6)),$$

with initial values (1.0, 1.0, 1.0), integration step 0.314 and accuracy 0.0001. The $x$ component was used in the above cases.

The time series from the "strange nonchaotic attractor" [18] was obtained by iterating the system:

$$\Theta_{n+1} = (\Theta_n + 2\pi\omega) mod(2\pi)$$

$$u_{n+1} = \Lambda(u_n \cos(\Theta) + v_n \sin(\Theta))$$

$$v_{n+1} = -0.5\Lambda(u_n \cos(\Theta) - v_n \cos(\Theta))$$

where $\omega = (5^{1/2} - 1)/2$ and $\Lambda = 2/(1 + u_n^2 + v_n^2)$. The $\Theta$ component was recorded.

# Captions

Tab. 1.: Interpretation table for all the eight possible results of the qualitative comparison combined with the quantitative tests using both the linear and nonlinear redundancy statistics.

Fig. 1.: (a): Linear redundancy $L^n(\tau)$, (b): redundancy $R^n(\tau)$, as functions of the lag $\tau$, computed from the time series generated by the linear autoregression. (c): $L^n(\tau)$, (d): $R^n(\tau)$ of the linear stochastic process constructed from a given spectrum (the surrogate data of the Rössler chaotic series).
The four different curves in each picture are the redundancies for different embedding dimensions, $n = 2 - 5$ reading from the bottom to the top. The redundancies $L^n(\tau)$ and $R^n(\tau)$ are plotted as $L^n(\tau)/(n-1)$ and $R^n(\tau)/(n-1)$, respectively.

Fig. 2.: (a): $L^n(\tau)$, (b): $R^n(\tau)$ of the Lorenz chaotic series, (c): $L^n(\tau)$, (d): $R^n(\tau)$ of the Rössler chaotic series.
The four different curves in each picture are the redundancies for different embedding dimensions, $n = 2 - 5$ reading from the bottom to the top. The redundancies $L^n(\tau)$ and $R^n(\tau)$ are plotted as $L^n(\tau)/(n-1)$ and $R^n(\tau)/(n-1)$, respectively.

Fig. 3.: (a): Linear redundancy statistic, (b): redundancy statistic for the linear stochastic process - the surrogate data of the Rössler chaotic series, 131,072 samples analyzed. (c): Linear redundancy statistic, (d): redundancy statistic for the same data, however, only 16,384-sample subset was analyzed.
The statistic is the difference between the value of a particular redundancy from the original data minus the mean value of a set of 30 surrogates, in number of standard deviations (SD) of the set of surrogates. The statistics are plotted as functions of the time lag $\tau$, the four different curves are the statistics for embedding dimensions $n = 2$ – full line, $n = 3$ – dash-and-dotted line, $n = 4$ – dashed line, and $n = 5$ – dotted line.

Fig. 4.: (a): $L^n(\tau)$, (b): $R^n(\tau)$ for the noisy sine series (50% of noise), (c): $L^n(\tau)$, (d): $R^n(\tau)$ for a surrogate of this noisy sine series. The four different curves in each picture are the redundancies for different embedding dimensions, $n = 2 - 5$ reading from the bottom to the top. The redundancies $L^n(\tau)$ and $R^n(\tau)$ are plotted as $L^n(\tau)/(n-1)$ and $R^n(\tau)/(n-1)$, respectively.

Fig. 5.: (a): Linear redundancy statistic, (b): redundancy statistic, (c): $L^n(\tau)$,

(d): $R^n(\tau)$ for the linear stochastic process - the surrogate data of the Rössler chaotic series, passed through the static exponential nonlinearity. The four different curves in each picture are the results for different embedding dimensions, $n = 2 - 5$, reading from the bottom to the top in the cases of the redundancies, in the cases of the statistics the different dimensions are distinguished by different line types — see the caption of Fig. 3.

Fig. 6.: (a): Linear redundancy statistic, (b): redundancy statistic for the sine series with 10% of additive Gaussian noise. The four different curves in each picture are the results for different embedding dimensions, $n = 2 - 5$, distinguished by different line types — see the caption of Fig. 3. (c): $L^2(\tau)$ for this series (full line) and the mean $L^2(\tau)$ of the set of its 30 surrogates (dashed line, the curves almost coincide), (d): $R^2(\tau)$ for this series (full line) and the mean $R^2(\tau)$ of the set of its 30 surrogates (dashed line). The time series length used was 131,072 samples.

Fig. 7.: Periodogram of the noisy (10%) sine series (65,536 frequency bins), the inset: the detail of the peak.

Fig. 8.: The same results as presented in Fig. 6, but for the series generated by the Rössler system in the periodic state. The time series length used was 131,072 samples.

Fig. 9.: (a): $L^2(\tau)$, (b): $R^2(\tau)$ for the linear stochastic process (the surrogate of the Rössler chaotic series, generated in 131,072-sample FFT) and its surrogates; (c): $L^2(\tau)$, (d): $R^2(\tau)$ for the Rössler periodic series and its surrogates. In both the cases the time series length used in the analysis was 16,384 samples. In each figure the redundancy from the original data is drawn by the full line, the dashed line represents the mean value of the redundancy for the set of 30 surrogates.

Fig. 10.: (a): $L^n(\tau)$, (b): $R^n(\tau)$, (c): redundancy statistic for the series generated by the system with the strange nonchaotic attractor.
The four different curves in each picture are the results for different embedding dimensions, $n = 2 - 5$, reading from the bottom to the top in the cases of the redundancies, in the case of the statistic the different dimensions are distinguished by different line types — see the caption of Fig. 3.
(d): $R^2(\tau)$ for this series (full line) and the mean $R^2(\tau)$ for the set of its 30 surrogates (dashed line).

Fig. 11.: (a): Linear redundancy statistic, (b): redundancy statistic, (c): mean

redundancy for the set of 30 surrogates, and (d): redundancy for the test data — the sunspots time series. Embedding dimension is $n = 2$. In plots c) and d) the dash-and-dotted line represents the mean value and the dashed lines the mean $\pm$ SD values of the redundancy of Gaussian random numbers, serving as the adjustment of numerical zero for these plots.

Fig. 12.: (a): Linear redundancy statistic, (b): redundancy statistic, (c): mean redundancy for the set of 30 surrogates, and (d): redundancy for the test data — the measles time series. Embedding dimension is $n = 2$. In plots c) and d) the dash-and-dotted line represents the mean value and the dashed lines the mean $\pm$ SD values of the redundancy of Gaussian random numbers, serving as the adjustment of numerical zero for these plots.

| Quantitative result in: | | Qualitative comparison: | |
| :---: | :---: | :---: | :---: |
| L-statistic | R-statistic | Similar | Different |
| non-significant | non-significant | (1) linear | (2) not observed |
| significant | non-significant | (3) not real | (4) not real |
| non-significant | significant | (5) nonlinear – trivial | (6) nonlinear |
| significant | significant | (7) linear or nonlinear – trivial | (8) nonlinear |